\pdfoutput=1
\documentclass[11pt,twoside,a4paper,cmspaper,final,collab]{cms-tdr}
\def\svnVersion{535e8bf-D}\def\svnDate{2021/06/21}\def\cmsCernNoTag{CERN-EP-2021-061}\def\cmsCernDate{\today}\def\cmsMessage{"Published in Physical Review D as \href{http://dx.doi.org/10.1103/PhysRevD.104.032013}{\doi{10.1103/PhysRevD.104.032013}.}"}
\begin{document}\cmsNoteHeader{HIG-20-009}

\newcommand{\XX}{\ensuremath{\textcolor{red}{XX}\xspace}}
\newcommand{\Htetmu}{\ensuremath{\PH\to\PGt_{\Pe}\PGt_{\PGm}}\xspace}
\newcommand{\Hteth}{\ensuremath{\PH\to\PGt_{\Pe}\tauh}\xspace}
\newcommand{\tauj}{\ensuremath{\PGt_{\text{j}}}\xspace}
\newcommand{\taue}{\ensuremath{\PGt_{\Pe}}\xspace}
\newcommand{\taum}{\ensuremath{\PGt_{\PGm}}\xspace}
\newcommand{\Ymutau}{\ensuremath{\abs{Y_{\PGm\PGt}}}\xspace}
\newcommand{\Ytaumu}{\ensuremath{\abs{Y_{\PGt\PGm}}}\xspace}
\newcommand{\Ytaue}{\ensuremath{\abs{Y_{\PGt\Pe}}}\xspace}
\newcommand{\Yetau}{\ensuremath{\abs{Y_{\Pe\PGt}}}\xspace}
\newcommand{\Htt}{\ensuremath{\PH\to\PGt\PGt}\xspace}
\newcommand{\HWW}{\ensuremath{\PH\to\PW\PW}\xspace}
\newcommand{\Hem}{\ensuremath{\PH\to\Pe\PGm}\xspace}
\newcommand{\Het}{\ensuremath{\PH\to\Pe\PGt}\xspace}
\newcommand{\Hmt}{\ensuremath{\PH\to\PGm\PGt}\xspace}
\newcommand{\BHem}{\ensuremath{\mathcal{B}(\PH\to\Pe\PGm)}\xspace}
\newcommand{\BHet}{\ensuremath{\mathcal{B}(\PH\to\Pe\PGt)}\xspace}
\newcommand{\BHmt}{\ensuremath{\mathcal{B}(\PH\to\PGm\PGt)}\xspace}
\newcommand{\BHtt}{\ensuremath{\mathcal{B}(\PH\to\PGt\PGt)}\xspace}
\newcommand{\BHWW}{\ensuremath{\mathcal{B}(\PH\to\PW\PW)}\xspace}
\newcommand{\Hmue}{\ensuremath{\PH\to\PGm\PGt_{\Pe}}\xspace}
\newcommand{\Hmumu}{\ensuremath{\PH\to\PGm\PGt_{\PGm}}\xspace}
\newcommand{\Hmuhad}{\ensuremath{\PH\to\PGm\tauh}\xspace}
\newcommand{\Hee}{\ensuremath{\PH\to\Pe\PGt_{\Pe}}\xspace}
\newcommand{\Hemu}{\ensuremath{\PH\to\Pe\PGt_{\PGm}}\xspace}
\newcommand{\Hehad}{\ensuremath{\PH\to\Pe\tauh}\xspace}
\newcommand{\muhad}{\ensuremath{\PGm\tauh}\xspace}
\newcommand{\mue}{\ensuremath{\PGm\PGt_{\Pe}}\xspace}
\newcommand{\mtm}{\ensuremath{\PGm\PGt_{\PGm}}\xspace}
\newcommand{\mutau}{\ensuremath{\PGm\PGt}\xspace}
\newcommand{\ehad}{\ensuremath{\Pe\tauh}\xspace}
\newcommand{\emu}{\ensuremath{\Pe\PGt_{\PGm}}\xspace}
\newcommand{\ete}{\ensuremath{\Pe\PGt_{\Pe}}\xspace}
\newcommand{\etau}{\ensuremath{\Pe\PGt}\xspace}
\newcommand{\egm}{\ensuremath{\Pe\PGm}\xspace}
\newcommand{\Zll}{\ensuremath{\PZ\to\ell\ell}\xspace}
\newcommand{\Zee}{\ensuremath{\PZ\to\Pe\Pe}\xspace}
\newcommand{\Zmm}{\ensuremath{\PZ\to\PGm\PGm}\xspace}
\newcommand{\Ztt}{\ensuremath{\PZ\to\PGt\PGt}\xspace}
\newcommand{\mcol}{\ensuremath{m_{\text{col}}}\xspace}
\newcommand{\mh}{\ensuremath{m_{\PH}}\xspace}
\newcommand{\msig}{\ensuremath{100\GeV<\mcol<150\GeV}\xspace}
\newcommand{\mjj}{\ensuremath{m_{\mathrm{jj}}}\xspace}
\newcommand{\mvis}{\ensuremath{m_{\text{vis}}}\xspace}
\newcommand{\wjets}{\ensuremath{\PW{+}\text{jets}}\xspace}
\newcommand{\zjets}{\ensuremath{\PZ{+}\text{jets}}\xspace}
\newcommand{\vectvis}{\ensuremath{\vec{\PGt}^{\text{vis}}}\xspace}
\newcommand{\tvis}{\ensuremath{\PGt^{\text{vis}}}\xspace}
\newcommand{\ptvis}{\ensuremath{\pt^{\vec{\PGt}^{\text{vis}}}}\xspace}
\newcommand{\ptnu}{\ensuremath{\pt^{\vec{\nu},\text{est}}}\xspace}
\newcommand{\dphitauhmet}{\ensuremath{\Delta\phi(\tauh,\ptvecmiss)}\xspace}
\newcommand{\dphimtauh}{\ensuremath{\Delta\phi(\PGm,\tauh)}\xspace}
\newcommand{\dphietauh}{\ensuremath{\Delta\phi(\Pe,\tauh)}\xspace}
\newcommand{\dphimmet}{\ensuremath{\Delta\phi(\PGm,\ptvecmiss)}\xspace}
\newcommand{\dphiemet}{\ensuremath{\Delta\phi(\Pe,\ptvecmiss)}\xspace}
\newcommand{\dphiem}{\ensuremath{\Delta\phi(\Pe,\PGm)}\xspace}
\newcommand{\detamtauh}{\ensuremath{\Delta\eta(\PGm,\tauh)}\xspace}
\newcommand{\detaetauh}{\ensuremath{\Delta\eta(\Pe,\tauh)}\xspace}
\newcommand{\mtmmet}{\ensuremath{\mT(\Pgm,\ptvecmiss)}\xspace}
\newcommand{\mtemet}{\ensuremath{\mT(\Pe,\ptvecmiss)}\xspace}
\newcommand{\mttmet}{\ensuremath{\mT(\Pgt,\ptvecmiss)}\xspace}
\newcommand{\irel}{\ensuremath{I^\ell_{\text{rel}}}\xspace}
\newcommand{\irele}{\ensuremath{I^\Pe_{\text{rel}}}\xspace}
\newcommand{\irelm}{\ensuremath{I^\PGm_{\text{rel}}}\xspace}
\newcommand{\aeta}{\ensuremath{\abs{\eta}}\xspace}
\newcommand{\dr}{\ensuremath{\Delta R}\xspace}
\providecommand{\cmsTable}[1]{\resizebox{\textwidth}{!}{#1}}

\cmsNoteHeader{HIG-20-009}

\title{Search for lepton-flavor violating decays of the Higgs boson in the \texorpdfstring{$\mutau$}{mutau} and \texorpdfstring{$\etau$}{etau} final states in proton-proton collisions at \texorpdfstring{$\sqrt{s} = 13\TeV$}{sqrt(s) = 13 TeV}}

\author*[inst1]{The CMS Collaboration}

\date{\today}

\abstract{
A search is presented for lepton-flavor violating decays of the Higgs boson to \mutau and \etau. The data set corresponds to an integrated luminosity of 137\fbinv collected at the LHC in proton-proton collisions at a center-of-mass energy of 13\TeV. No significant excess has been found, and the results are interpreted in terms of upper limits on lepton-flavor violating branching fractions of the Higgs boson. The observed (expected) upper limits on the branching fractions are, respectively, $\BHmt < 0.15 \, (0.15)\%$ and $\BHet < 0.22 \, (0.16)\%$ at 95\% confidence level.
}

\hypersetup{%
pdfauthor={CMS Collaboration},%
pdftitle={Search for lepton flavor violating decays of the 125 GeV Higgs boson to mu-tau and e-tau in proton-proton collisions at sqrt s = 13 TeV},%
pdfsubject={CMS},%
pdfkeywords={CMS, lepton flavor violation, higgs}}

\maketitle

\section{Introduction}

One of the main goals of the LHC program is to search for processes beyond the standard model (BSM). The properties and decays of the Higgs boson (\PH) are thus far consistent with expectations of the standard model (SM)~\cite{Aad:2015gba, Khachatryan:2014jba, Chatrchyan:2012jja, Aad:2013xqa, Khachatryan:2014kca, Sirunyan:2017exp}. However, there is considerable motivation to search for BSM decays of the Higgs boson. The lepton-flavor violating (LFV) decays of the Higgs boson~\cite{Aad:2012tfa, Chatrchyan:2012ufa, Chatrchyan:2013lba} can provide possible signatures of such processes. A previous investigation of the combined results from the CMS experiment constrained the branching fraction for $\mathcal{B}(\PH\to\mathrm{BSM})$ to $<$0.36 at the 95\% confidence level (\CL), leaving the possibility for a large contribution for these decays~\cite{Sirunyan:2018koj}.

The LFV decays \Hem, \Het, or \Hmt are forbidden in the SM, but take place through the LFV Yukawa couplings $Y_{\egm}$, $Y_{\etau}$, or $Y_{\mutau}$, respectively~\cite{Harnik:2012pb}. The LFV decays arise in models with more than one Higgs boson doublet~\cite{Bjorken:1977vt}, certain supersymmetric models~\cite{DiazCruz:1999xe, Han:2000jz, Arhrib:2012ax}, composite Higgs models~\cite{Agashe:2009di, Azatov:2009na}, models with flavor symmetries~\cite{Ishimori:2010au}, the Randall--Sundrum model of extra spatial dimensions~\cite{Perez:2008ee, Casagrande:2008hr, Buras:2009ka, Blanke:2008zb, Albrecht:2009xr}, and other models~\cite{Giudice:2008uua, AguilarSaavedra:2009mx, Goudelis:2011un, McKeen:2012av, Pilaftsis:1992st, Korner:1992zk}.

Here we report a search for LFV decays of the Higgs boson in the \mutau and \etau channels performed using data collected by the CMS experiment in proton-proton (\Pp{}\Pp) collisions at a center-of-mass energy of 13\TeV during the 2016--2018 data-taking period, corresponding to an integrated luminosity of 137\fbinv. The CMS experiment set upper limits of 0.25\% and 0.61\%~\cite{Sirunyan:2017xzt} and the ATLAS experiment set upper limits of 0.28\% and 0.47\%~\cite{Aad:2019ugc} on \BHmt and \BHet at 95\% \CL, respectively, based on the 2016 data set, corresponding to an integrated luminosity of 36\fbinv.

The presence of an LFV Higgs boson coupling leads to processes such as $\PGm\to\Pe$, $\PGt\to\PGm$, and $\PGt\to\Pe$ to proceed via a virtual Higgs boson~\cite{Shanker:1981mj, McWilliams:1980kj}. The experimental limits on these decays yield indirect constraints on \BHem, \BHmt, and \BHet~\cite{Harnik:2012pb, Blankenburg:2012ex}. The null result for $\PGm\to\Pe\Pgg$~\cite{Adam:2013mnn} strongly constrains \BHem to $<10^{-8}$. Searches for rare \PGt lepton decays~\cite{Celis:2013xja}, such as $\PGt\to\Pe\Pgg$ and $\PGt\to\PGm\Pgg$, and the measurement of the electron and muon magnetic moments, have set constraints on \BHet and \BHmt of ${\approx}10\%$, which are much less stringent than those from the direct searches.

Our search is performed in the \muhad, \mue, \ehad, and \emu channels, where \tauh, \taue, and \taum correspond to the $\PGt\to \text{hadrons}$, electron, and muon decay channels of \PGt leptons, respectively, each accompanied by its corresponding neutrinos. The \ete and \mtm decays are not considered because of the large background contribution from \PZ/$\gamma^{*}$ decays.

Our search significantly improves the sensitivity relative to similar previous studies~\cite{Khachatryan:2016rke, Aad:2019ugc, Sirunyan:2017xzt}. The search makes use of boosted decision tree (BDT) discriminants to distinguish signal from background in the distributions which are then used for performing the statistical analysis. Constraints on the branching fractions are extracted under the assumption that only one of the LFV decays contributes additionally to the SM Higgs boson total width. The constraints on the branching fractions are correspondingly translated into limits on the $Y_{\etau}$ and $Y_{\mutau}$ LFV Yukawa couplings.

This paper is organized as follows: a description of the CMS detector is given in Section~\ref{sec:detector}, collision data and simulated events are discussed in Section~\ref{sec:dataset}, event reconstruction is described in Section~\ref{sec:event_reco}, and event selection is described separately for the four decay channels in Section~\ref{sec:selection}. Background estimation and systematic uncertainties are described in Sections~\ref{sec:bkgEstimation} and~\ref{sec:sysUnc}, respectively. Results are presented in Section~\ref{sec:results} and the paper is summarized in Section~\ref{sec:summary}.

\section{The CMS detector}
\label{sec:detector}

The CMS detector consists of a silicon pixel and strip tracker, a lead tungstate crystal electromagnetic calorimeter (ECAL), a brass and scintillator hadron calorimeter (HCAL), and a muon system composed of gaseous detectors. Each subdetector consists of a barrel and two endcap sections. The central feature of the CMS detector is a superconducting solenoid of 6\unit{m} internal diameter, providing a magnetic field of 3.8\unit{T}. The tracking systems and the calorimeters are contained within the solenoid volume; the muon chambers are embedded in the steel flux-return yoke outside the solenoid. Forward calorimeters extend the pseudorapidity ($\eta$) coverage provided by the barrel and endcap detectors.

Events of interest are selected using a two-tiered trigger system. The first level, composed of custom hardware processors, uses information from the calorimeters and muon detectors to select events at a rate of ${\approx}100\unit{kHz}$ within a fixed latency of ${\approx}4\mus$~\cite{Sirunyan:2020zal}. The second level, the high-level trigger, consists of a farm of processors running a version of the full event reconstruction software optimized for fast processing that reduces the event rate to ${\approx}1\unit{kHz}$ before data storage~\cite{Khachatryan:2016bia}. A more detailed description of the CMS detector, together with a definition of the coordinate system and kinematic variables, can be found in Ref.~\cite{Chatrchyan:2008zzk}.

\section{Collision data and simulated events}
\label{sec:dataset}

The search presented makes use of \Pp{}\Pp collisions collected at the CMS experiment at a center-of-mass energy of 13\TeV in 2016--2018. The total integrated luminosity amounted to 35.9\fbinv in 2016, 41.5\fbinv in 2017, and 59.7\fbinv in 2018. Single-muon triggers with isolation criteria are used to collect the data in the \muhad channel. Electron-muon triggers are used to collect data in the \mue and \emu channels. Triggers requiring a single isolated electron, or a combination of an electron and \tauh, are used in the \ehad channel. The trigger thresholds are mentioned in Section~\ref{sec:selection}.

Simulated events are used to model signal and background events using several event generators. In all cases parton showering, hadronization, and underlying event properties are modeled using \PYTHIA~\cite{Sjostrand:2014zea} version 8.212. The \PYTHIA parameters affecting the description of the underlying event are set to the CUETP8M1 tune in 2016~\cite{Khachatryan:2015pea}, except for the \ttbar events that use the CP5 tune which is used for all the events in 2017 and 2018~\cite{Sirunyan:2019dfx}. The NNPDF3.0 parton distribution functions (PDFs) for all 2016 events and the NNPDF3.1 PDFs for the 2017 and 2018 events~\cite{Ball:2017nwa}.

The simulation of interactions in the CMS detector is based on \GEANTfour~\cite{Agostinelli:2002hh}, using the same reconstruction algorithms as used for data. The Higgs bosons are generated in \Pp{}\Pp collisions predominantly through gluon fusion (\Pg{}\Pg{}\PH)~\cite{Georgi:1977gs}, but also via vector boson fusion (VBF)~\cite{Cahn:1986zv}, and in association with a vector boson (\PW or \PZ)~\cite{Glashow:1978ab}. Such events are generated at next-to-leading order (NLO) in perturbative quantum chromodynamics (QCD) with the \POWHEG v2.0 generator~\cite{Nason:2004rx, Frixione:2007vw, Alioli:2010xd, Alioli:2010xa, Alioli:2008tz, Bagnaschi:2011tu}, using the implementation of Refs.~\cite{Heinrich:2017kxx, Buchalla:2018yce}. For the LFV signal, we consider just the Higgs bosons via the \Pg{}\Pg{}\PH and VBF mechanisms as the contribution from associated vector boson production is found to be negligible.

The \Ztt background events are estimated in a data-driven manner using the embedding technique because it provides a better description of jets, pileup, as well as detector noise and resolution effects compared to simulation. These events are obtained from data with well identified \Zmm decays from which muons are removed, and simulated \PGt leptons are embedded with the same kinematic variables as the replaced muons. The \MGvATNLO generator~\cite{Alwall:2014hca} (version 2.2.2 in 2016, version 2.4.2 in 2017 and 2018) is used to simulate the $\Zee{+}\text{jets}$ and $\Zmm{+}\text{jets}$ processes, the \wjets background process, and the electroweak (EW) \PW/\PZ events. They are simulated at leading order with the MLM jet matching and merging schemes~\cite{Alwall:2007fs}.

Diboson production is simulated at NLO using the \MGvATNLO generator with the FxFx jet-matching and merging scheme~\cite{Frederix:2012ps}. Top quark-antiquark pair and single top quark production are generated at NLO using \POWHEG.

The effect of pileup, where events of interest have multiple \Pp{}\Pp interactions in the same bunch crossing, is taken into account in simulated events by generating concurrent minimum bias events. All simulated events are weighted to match the pileup distribution observed in the data.

\section{Event reconstruction}
\label{sec:event_reco}

The particle flow (PF) algorithm~\cite{CMS-PRF-14-001} reconstructs and identifies each particle in an event through an optimized combination of information from the various subdetectors of the CMS detector. In this process, identifying the PF candidate type (photons, electrons, muons, charged, and neutral hadrons) plays an important role in determining particle direction and energy. The candidate vertex with the largest value of summed physics object $\pt^2$, where \pt is the transverse momentum, is taken to be the primary \Pp{}\Pp interaction vertex (PV). The physics objects are returned by a jet finding algorithm~\cite{Cacciari:2008gp, Cacciari:2011ma} applied to all charged tracks associated with the vertex, plus the corresponding associated missing transverse momentum (\ptvecmiss).

An electron is identified as a track from the PV combined with one or more ECAL energy clusters. These clusters correspond to the electron and possible bremsstrahlung photons emitted when passing through the tracker. Electrons are accepted in the range  $\aeta<2.5$, except for the region $1.44<\aeta<1.57$ where the detector's service infrastructure is located. They are identified with an efficiency of 80\% using a multivariate discriminator that combines observables sensitive to the amount of bremsstrahlung energy deposited along the electron trajectory, the geometric and momentum matching between the electron trajectory and associated clusters, and the distribution in shower energy in the calorimeters~\cite{Sirunyan:2020ycc}. Electrons from photon conversions are removed. The electron momentum is estimated by combining the energy measurement in the ECAL with the momentum measurement in the tracker. The momentum resolution for electrons with $\pt\approx45\GeV$ from $\PZ\to\Pe\Pe$ decays ranges from 1.7 to 4.5\% depending on the \aeta. It is generally better in the barrel region than in the endcaps~\cite{Khachatryan:2015hwa}.

Muons are measured in the $\aeta<2.4$ range using the drift tube, cathode strip chamber, and resistive plate chamber technologies. The efficiency to reconstruct and identify muons is greater than 96\%. Matching muons to tracks measured in the silicon tracker results in a relative \pt resolution for muons with \pt up to 100\GeV of 1\% in the barrel and 3\% in the endcaps~\cite{Sirunyan:2018fpa}.

The muon or electron isolations are measured relative to its $\pt^{\ell}$, where $\ell$ is either \PGm or \Pe, values by summing over the scalar \pt of PF particles in a cone of $\dr=0.4$ or 0.3 around the lepton:
\ifthenelse{\boolean{cms@external}}{
  \begin{multline*}
    \irel = \Bigg( \sum \pt^{\text{PV\,charged}} + \text{max}\Big[0, \sum \pt^{\text{neutral}} \\
          + \sum \pt^\Pgg - \pt^{\mathrm{PU}}(\ell) \Big] \Bigg) \Big/ \pt^\ell,
  \end{multline*}
  }{
  \begin{linenomath*}
    \begin{equation*}
      \irel = \left( \sum \pt^{\text{PV\,charged}} + \text{max}\left[0, \sum \pt^{\text{neutral}} + \sum \pt^\Pgg - \pt^{\mathrm{PU}}(\ell)\right]\right) \biggm/ \pt^\ell,
    \end{equation*}
  \end{linenomath*}
}
where $\pt^{\text{PV\,charged}}$, $\pt^{\text{neutral}}$, and $\pt^\Pgg$ indicate the \pt of a charged hadron, a neutral hadron, and a photon within the cone, respectively. The neutral particle contribution to isolation from pileup, $\pt^\mathrm{PU}(\ell)$, is estimated from the area of jet and its median energy density in the event~\cite{Cacciari:2008gn} for the electron. For the muon, half of the \pt sum of charged hadrons within the isolation cone, not originating from the PV, is used instead. The charged-particle contribution to isolation from the pileup is rejected by requiring the tracks to originate from the PV.

The reconstruction of \tauh is performed using the hadrons-plus-strips algorithm, which combines the signature for charged hadrons composed of tracks left in the tracker and energy depositions in the calorimeters with the signature for electrons or photons from neutral pion decays that are reconstructed as electromagnetic ``strips'' in $\eta$-$\phi$ space~\cite{Sirunyan:2018pgf}, where $\phi$ is the azimuth in radians. The combination of these signatures provides the four-vector for the parent \tauh. Based on the overall neutral versus charged contents of the \tauh reconstruction, a decay mode is assigned as $\mathrm{h}^{\pm}$, $\mathrm{h}^{\pm}\pi^{0}$, $\mathrm{h}^{\pm}\mathrm{h}^{\mp}\mathrm{h}^{\pm}$, or $\mathrm{h}^{\pm}\mathrm{h}^{\mp}\mathrm{h}^{\pm}\pi^{0}$, where $\mathrm{h}^{\pm}$ denotes a charged hadron. It has a reconstruction efficiency of ${\approx}80\%$.

The \tauh reconstructed using the hadrons-plus-strips algorithm must be well identified to reject jets, muons, and electrons misidentified as \tauh. A deep neural network (DNN) discriminator is used to further improve \tauh identification~\cite{CMS-DP-2019-033}. The input variables to the DNN include \tauh lifetime, isolation, and information of PF candidates reconstructed within the \PGt lepton signal or isolation cones. A \pt dependent threshold on the output of the DNN is used to distinguish \tauh from jets. The chosen working point (WP) has a \tauh identification efficiency of 70\% with a misidentification probability of 1\%. The DNN can reject electrons and muons misidentified as \tauh using dedicated criteria based on the consistency between the tracker, calorimeter, and muon detector measurements. In the \muhad or \ehad channel, we use a WP that has an efficiency of 97.5\% or 87.5\% with a misidentification probability of 1--2\% or 0.2--0.3\% to discriminate \tauh against electrons, and we use a WP that has an efficiency of 99.6\% or 99.8\% with a misidentification probability of 0.04\% or 0.06\% to discriminate \tauh against muons, respectively.

Charged hadrons are defined as PF tracks from the PV not reconstructed as electrons, muons, or \tauh leptons. Neutral hadrons are identified as HCAL energy clusters not assigned to any charged hadron or as excesses in ECAL or HCAL energies relative to the small charged-hadron energy deposit. All the PF hadron candidates are clustered into jets using the infrared- and collinear-safe anti-\kt algorithm~\cite{Cacciari:2008gp} with a distance parameter of 0.4. Jet momentum is determined as the vectorial sum of all particle momenta in the jet. It is found from simulation to be, on average, within 5--10\% of the true momentum over the entire \pt spectrum and detector acceptance~\cite{CMS-PAS-JME-16-003}. Jets that contain {\cPqb} quarks are tagged using a DNN-based algorithm, using a WP with efficiency of 70\% for a misidentification probability for light-flavor jets of 1\%~\cite{Sirunyan:2017ezt}.

The interactions from pileup add more tracks and calorimetric energy depositions, thereby increasing the apparent jet momenta. To mitigate this effect, tracks identified as originating from pileup vertices are discarded, and an offset correction is applied to correct the remaining contributions~\cite{Cacciari:2007fd}. Jet energy corrections are obtained from simulation studies so that the average measured energy of jets matches that of particle level jets. In-situ measurements of the momentum balance in photon+jet, \zjets, and multijet events are used to determine any residual differences between the jet energy scale in data and simulation, and appropriate corrections are applied~\cite{Khachatryan:2016kdb}. Additional selection criteria are applied to each jet to remove jets potentially dominated by instrumental effects or reconstruction failures. When combining information from the entire detector, the jet energy resolution typically amounts to 15\% at 10\GeV, 8\% at 100\GeV, and 4\% at 1\TeV. The variable $\dr=\sqrt{\smash[b]{(\Delta\eta)^2+(\Delta\phi)^2}}$ is used to measure the separation between reconstructed objects in the detector. Any jet within $\dr=0.5$ of identified leptons is removed. The reconstructed jets must have a $\pt>30\GeV$ and $\aeta<4.7$. Data collected in the high \aeta region of the ECAL endcaps were affected by noise during the 2017 data taking. This is mitigated by discarding events containing jets with $\pt<50\GeV$ and $2.65<\aeta<3.14$ in the 2017 data.

The vector \ptvecmiss is computed as the negative of the vector \pt sum of all the PF candidates in an event, and its magnitude is denoted as \ptmiss~\cite{Sirunyan:2019kia}. The \ptvecmiss is modified to account for corrections to the reconstructed jets' energy scale in the event. Anomalous high-\ptmiss events can originate from various reconstruction failures, detector malfunctions, or backgrounds not from beam-beam sources. Such events are rejected using event filters designed to identify more than 85--90\% of the spurious high-\ptmiss events with a mistag rate of less than 0.1\%~\cite{Sirunyan:2019kia}. In addition to the event-filtering algorithms, we require the jets to have a neutral hadron energy fraction smaller than 0.9, which rejects more than 99\% of jets due to detector noise, independent of jet \pt, with a negligible mistag rate. Corrections applied to the \ptvecmiss reduce the mismodeling of \ptvecmiss in simulated \PZ, \PW, and Higgs boson events. The corrections are applied to simulated events based on the vectorial difference in the measured \ptvecmiss and total \pt of neutrinos originating from the decay of the \PZ, \PW, or Higgs bosons. Their average effect is the reduction of the magnitude of the \ptvecmiss obtained from the simulation by a few \GeV.

\section{Event selection}
\label{sec:selection}

The signal topology consists of a muon or an electron and an oppositely charged \PGt lepton. The events in the \mutau and \etau channels are further divided into leptonic and hadronic channels based on the \PGt lepton decay mode (\taum, \taue, or \tauh). Jets misidentified as electrons or muons are suppressed by imposing isolation requirements described above. A set of loose selection criteria, known as the 'preselection', is first defined in each channel's respective signature. Events with more than two jets are not considered in the search. Each channel's events are then divided into categories based on the number of jets in the event (0-, 1-, or 2-jet) to enhance different Higgs boson production mechanisms. The dominant production mechanism contributing to the signal yield in the 0-jet category is ggH, while in the 1-jet category, it is ggH with initial-state radiation. The 2-jet category is further split into two based on the invariant mass of the two jets (\mjj). The optimization resulted in a threshold of $550\GeV$ and $500\GeV$ on \mjj for the \mutau and \etau channels, respectively, for the sensitivity optimization. The dominant production mechanism is ggH with initial-state radiation for events with $\mjj<550\GeV$ and $<500\GeV$, while it is VBF for events with $\mjj>550\GeV$ and $>500\GeV$ for the \mutau and \etau channels, respectively.

\begin{figure*}[htbp]
  \begin{center}
  \includegraphics[width=0.45\textwidth]{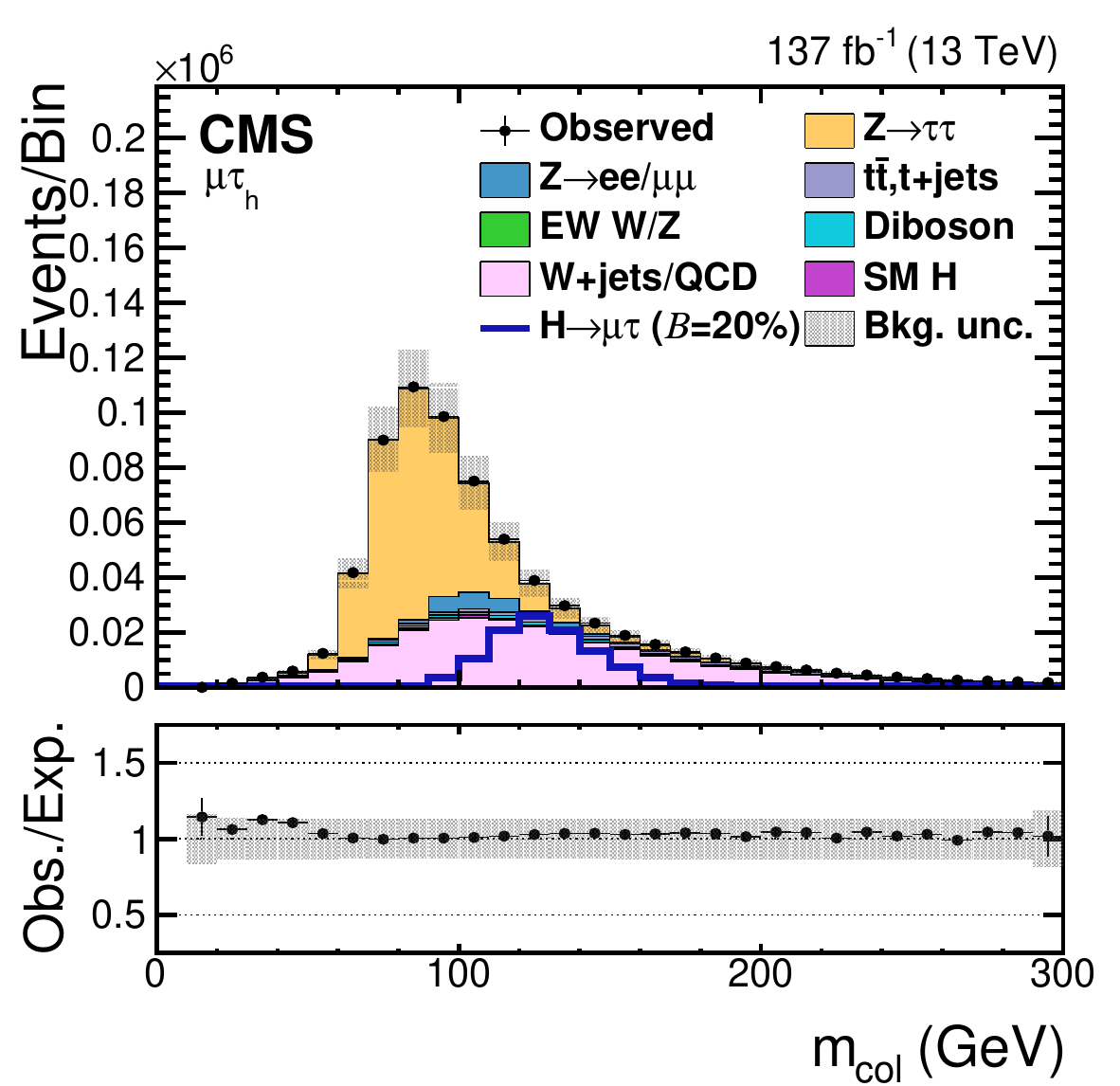}
  \includegraphics[width=0.45\textwidth]{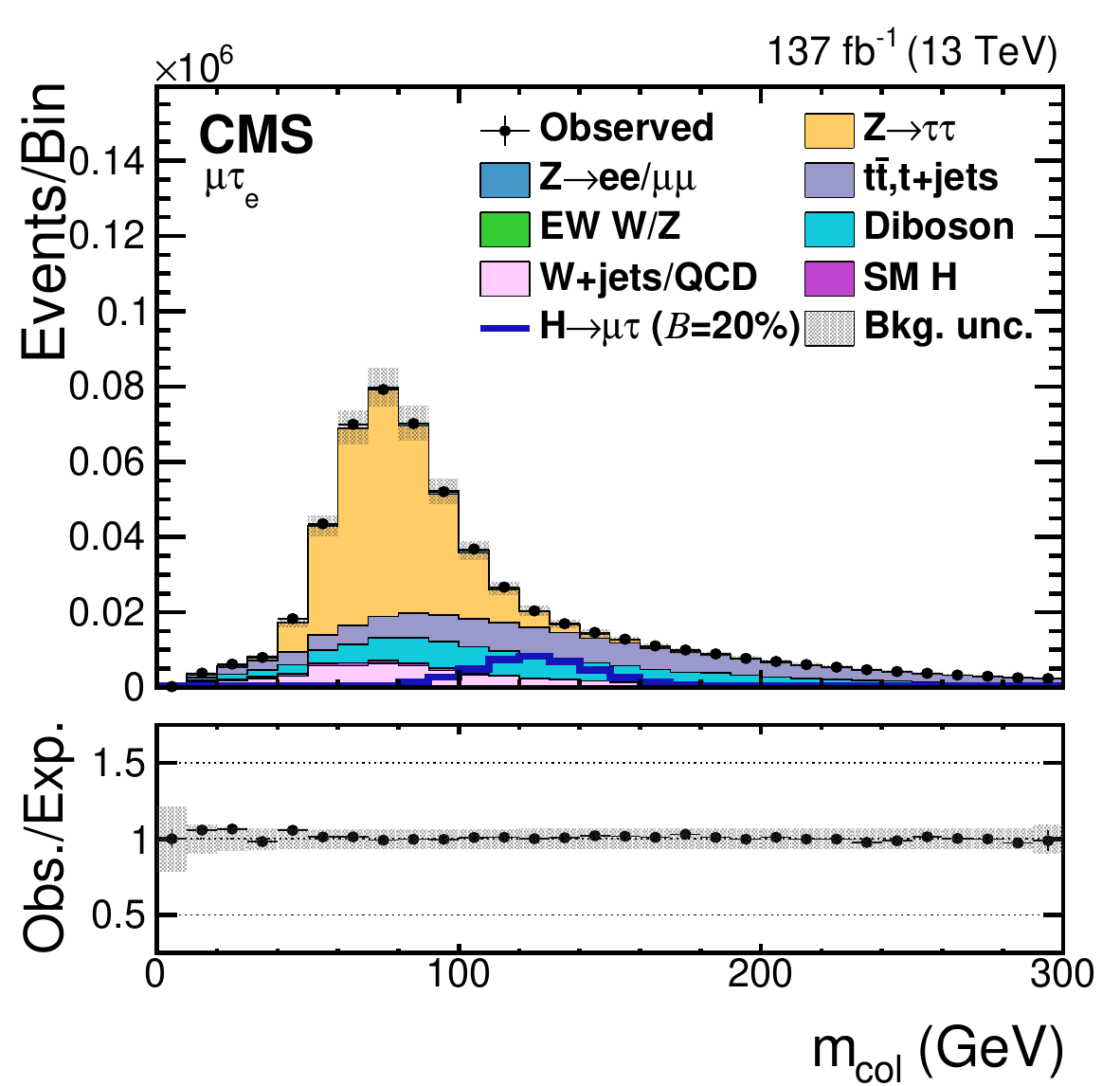}\\
  \includegraphics[width=0.45\textwidth]{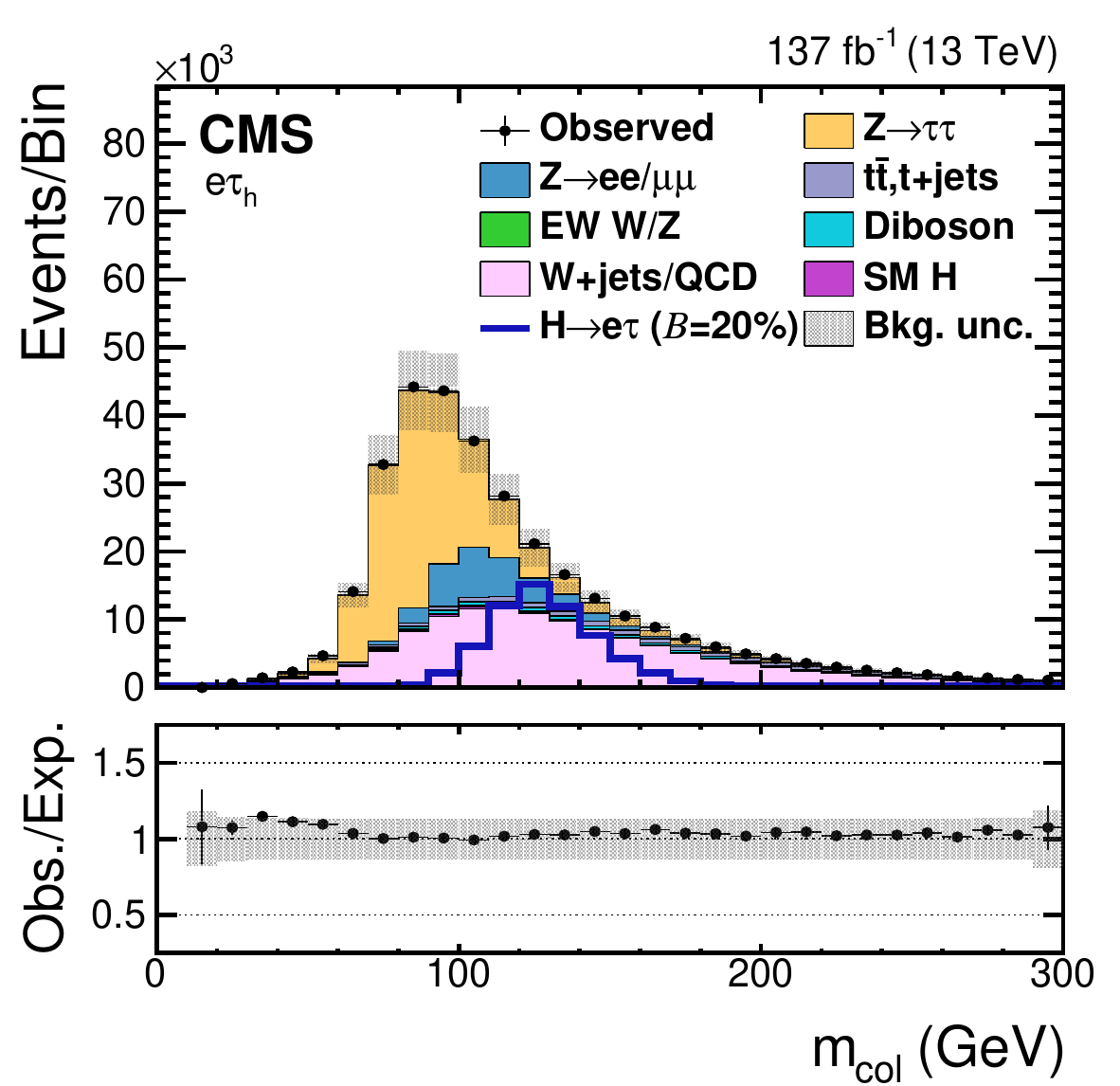}
  \includegraphics[width=0.45\textwidth]{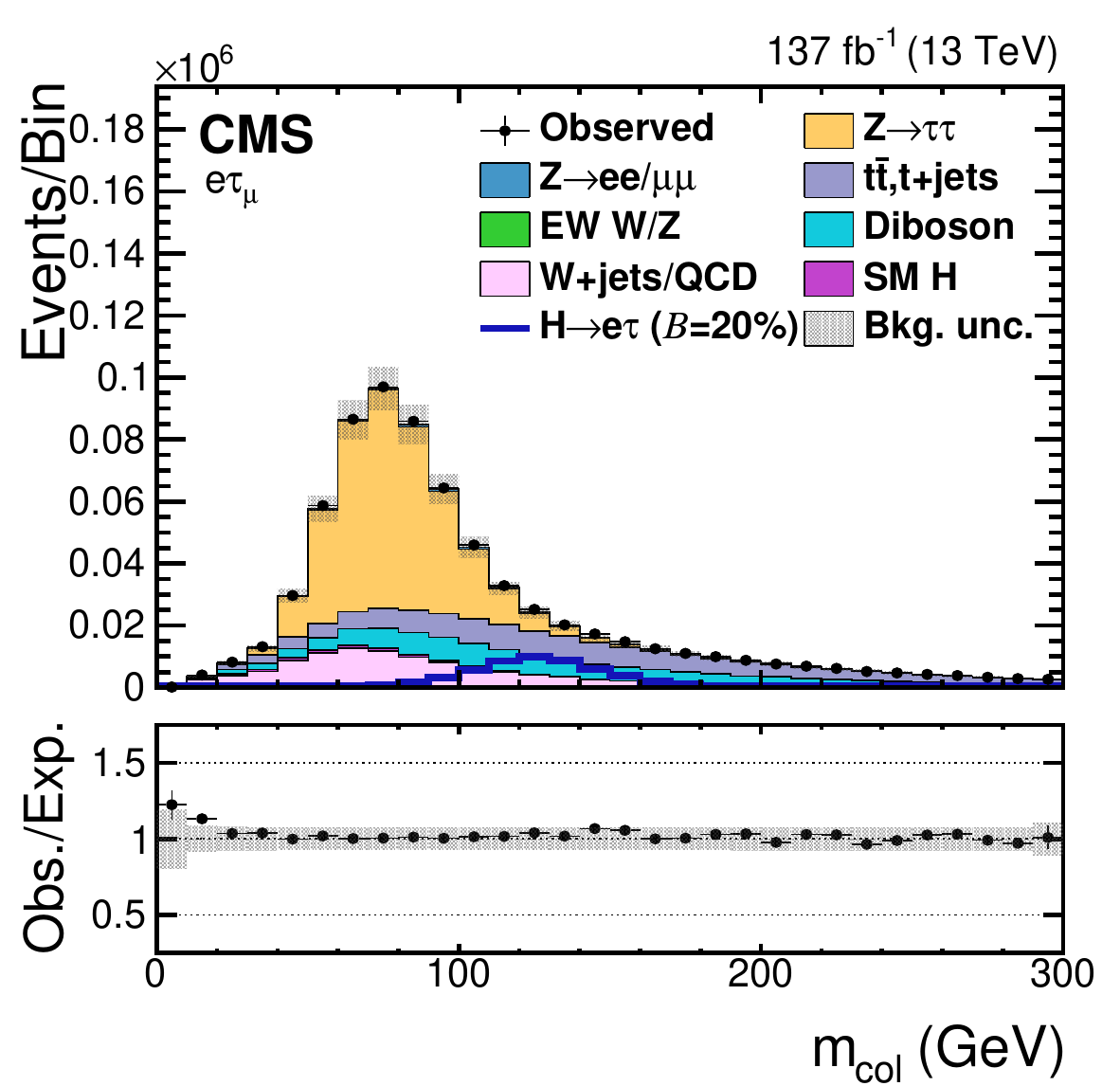}\\
  \caption{Collinear mass distributions for the data and background processes. A $\BHmt=20\%$ and $\BHet=20\%$ are assumed for the two signal processes. The channels are \Hmuhad (upper row left), \Hmue (upper row right), \Hehad (lower row left), and \Hemu (lower row right). The lower panel in each plot shows the ratio of data and estimated background. The uncertainty band corresponds to the background uncertainty in which the statistical and systematic uncertainties are added in quadrature.}
  \label{fig:coll_mass}
  \end{center}
\end{figure*}

A variable providing an estimate of \mh using the observed decay products of the Higgs boson, the collinear mass, is defined as $\mcol=\mvis/\sqrt{\smash[b]{x_\tvis}}$, where \mvis is the visible invariant mass of the \PGt-\PGm or \PGt-\Pe system and $x_\tvis$ is the fraction of the \PGt lepton \pt carried by the visible decay products of the \PGt lepton (\vectvis). The definition is based on the ``collinear approximation'' with the observation that, since $\mh \gg m_\PGt$, the \PGt lepton decay products are Lorentz-boosted in the direction of the \PGt lepton~\cite{Ellis:1987xu}. The momentum of neutrino(s) from the \PGt lepton decay can be approximated to have the same direction as the \vectvis. The component of the \ptvecmiss in the direction of the \vectvis is used to estimate the transverse component of the neutrino momentum (\ptnu). This information is combined to estimate the $x_\tvis$ which is defined as $\smash[b]{x_\tvis=\ptvis/(\ptvis+\ptnu)}$. The collinear mass distributions of simulated signal, data, and backgrounds in each channel are shown in Fig.~\ref{fig:coll_mass}.

The transverse mass $\mT(\ell)$ is a variable constructed from the lepton \pt and the \ptvecmiss vectors: $\mT(\ell)=\sqrt{\smash[b]{2\abs{\ptvec^\ell}\abs{\ptvecmiss}(1-\text{cos}\Delta\phi_{\ell, \ptvecmiss})}}$, where $\Delta\phi_{\ell, \ptvecmiss}$ is the angle in the transverse plane between the lepton and the \ptvecmiss, used to discriminate the Higgs boson signal from the \wjets background. The $\mT(\ell)$ distribution for the signal defined using visible decay products of the \PGt lepton peaks at lower values, while it peaks at higher values for the \wjets background.

To improve discrimination between signal and background events, a BDT is trained using the \textsc{tmva} toolkit of the \textsc{Root} analysis package~\cite{Hocker:2007ht}. A BDT is trained in each channel using a mixture of simulated signal events comprising the ggH and VBF processes, weighted according to their expected yield from SM production cross sections. In hadronic channels, the dominant sources of background come from the \Ztt process and events with misidentified leptons. The background used for training a BDT in the hadronic channels is obtained from data containing misidentified lepton events of the same electric charge for both the leptons and \Zll ($\ell=\Pe,\PGm,\PGt$) simulated events with their applied signal selections. In leptonic channels, the dominant sources of background come from the \Ztt process, the \ttbar process, and events with misidentified leptons. The background used for training a BDT in the leptonic channels is obtained from \ttbar and \Zll simulated events mixed and weighted according to their expected yield from SM production cross sections. Additional background for training comes from events with misidentified leptons in a control region (CR) in data, where the isolation requirements are inverted with the same electric charge for both the leptons. A detailed description of the different background processes and their estimation is given in Section~\ref{sec:bkgEstimation}.

The input variables to the BDT are mentioned separately for each channel below. The input variables are chosen based on their separation power as observed during training the BDT. The trained BDT is validated in a dedicated background enriched validation region (VR) for each channel and is detailed in Section~\ref{sec:bkgEstimation}. In all the channels, events containing additional electrons, muons, or \tauh candidates are vetoed. Also, events with at least one {\cPqb}-tagged jet are rejected to suppress the \ttbar background. After applying the selections, a maximum likelihood fit is performed to the BDT discriminant distributions in each channel. The various systematic uncertainties are incorporated as nuisance parameters in the fit. The BDT discriminant distributions in all the channels are shown after determining the best fit values of the nuisance parameters from the fit to the signal-plus-background hypothesis, as discussed later in Section~\ref{sec:sysUnc}.

\subsection{\texorpdfstring{\ensuremath{\PH \to \mu \tauh}\xspace}{Hmutauh}}

In this channel, the preselection requires a muon and \tauh of opposite electric charge with a separation of $\dr>0.5$. The trigger requires the presence of an isolated muon with a \pt threshold of 24\GeV. In 2017, this trigger is ``prescaled'', which means that only a fraction of events selected will pass the trigger. Hence, it is used in conjunction with another trigger based on the presence of an isolated muon with a \pt threshold of 27\GeV. The muon is required to have $\pt>26\GeV$, $\aeta<2.1$, and $\irelm<0.15$. The \tauh is required to have $\pt>30\GeV$ and $\aeta<2.3$. The selections for the \muhad channel are summarized in Table~\ref{tab:evtselection_mutau}.

\begin{figure*}[htbp]
  \begin{center}
  \includegraphics[width=0.45\textwidth]{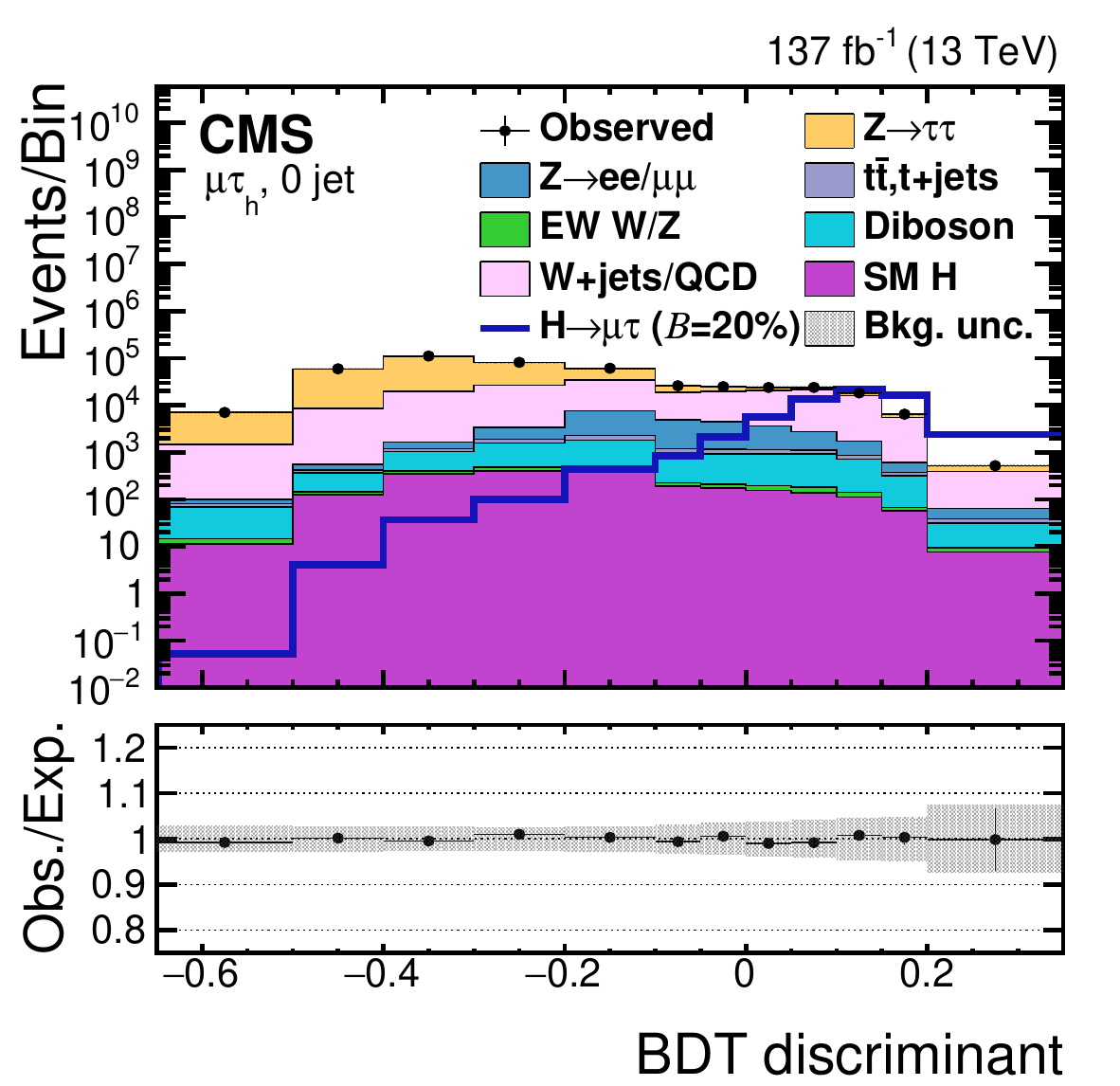}
  \includegraphics[width=0.45\textwidth]{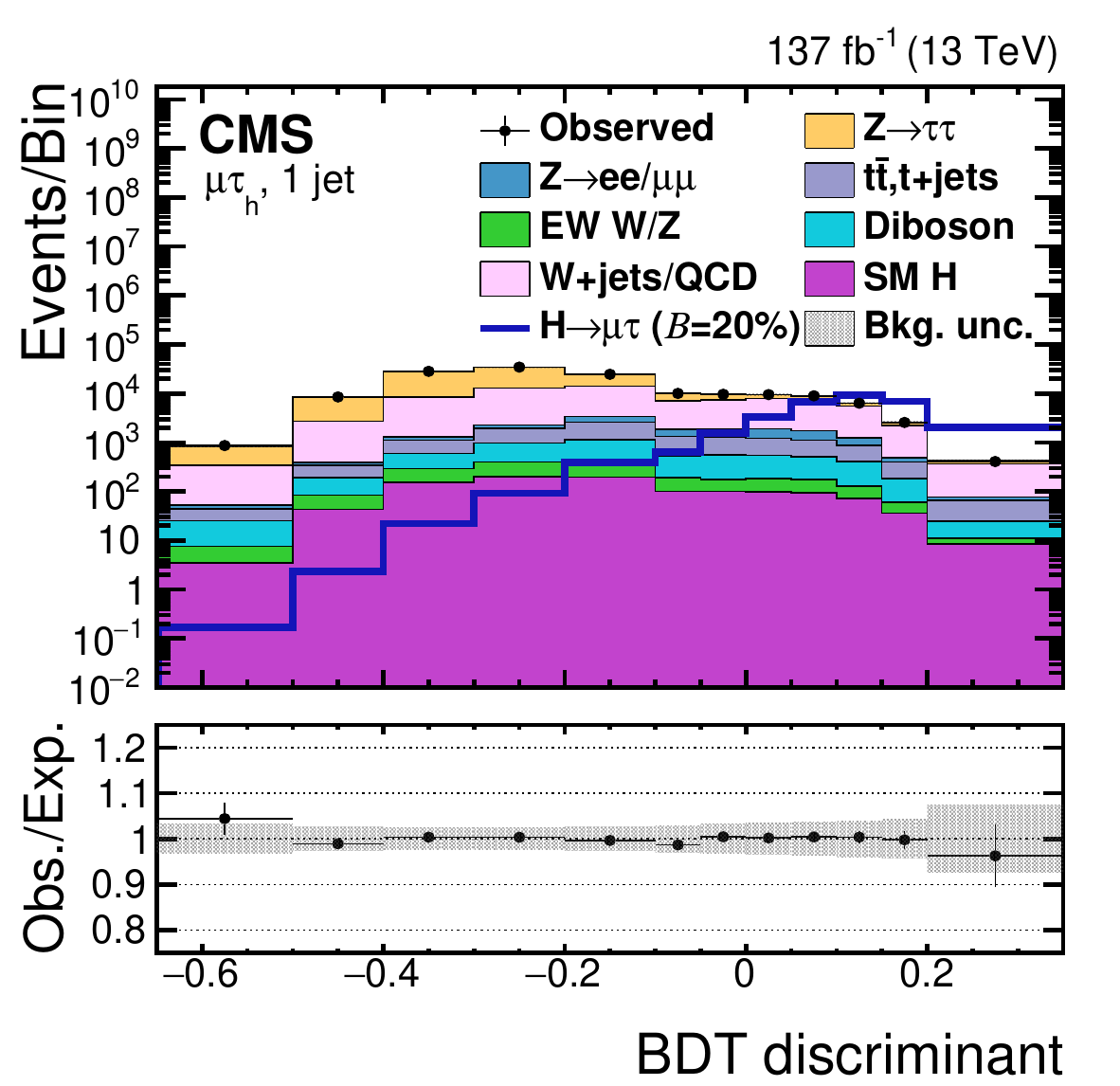}\\
  \includegraphics[width=0.45\textwidth]{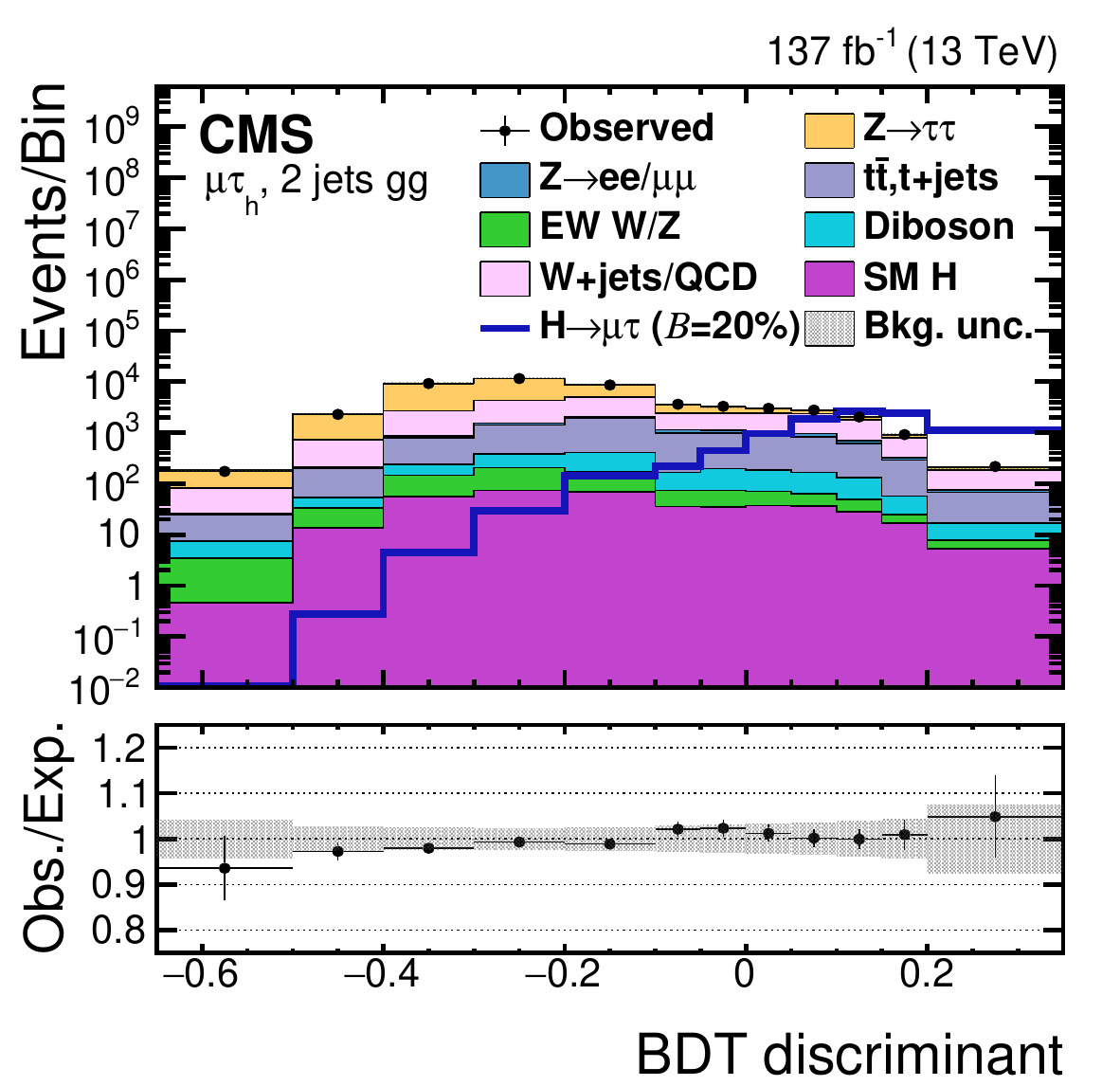}
  \includegraphics[width=0.45\textwidth]{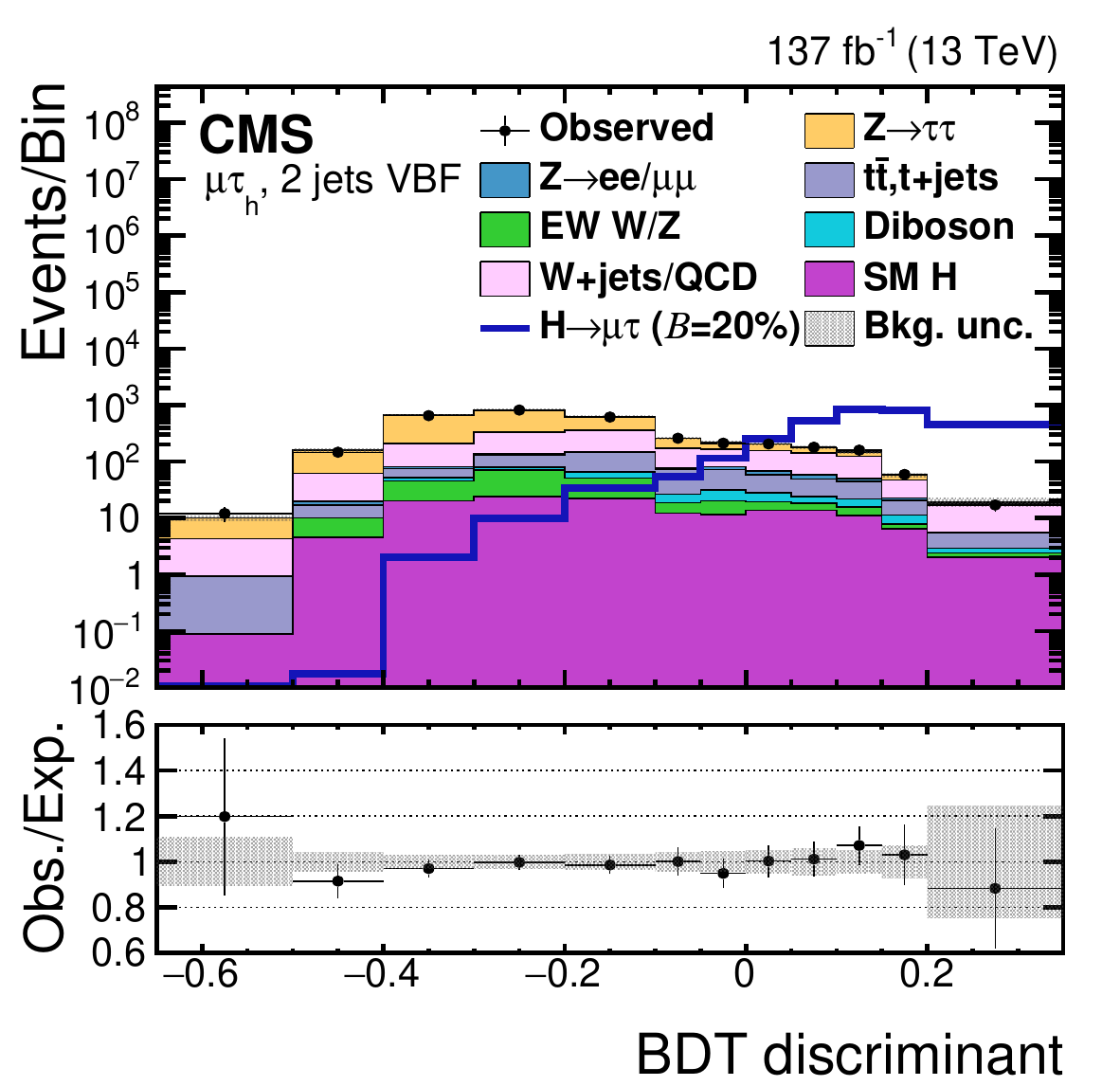}\\
  \caption{BDT discriminant distributions for the data and background processes in the \Hmuhad channel. A $\BHmt=20\%$ is assumed for the signal. The channel categories are 0 jets (upper row left), 1 jet (upper row right), 2 jets \Pg{}\Pg{}\PH (lower row left), and 2 jets VBF (lower row right). The lower panel in each plot shows the ratio of data and estimated background. The uncertainty band corresponds to the background uncertainty in which the post-fit statistical and systematic uncertainties are added in quadrature.}
  \label{fig:bdt_muhad}
  \end{center}
\end{figure*}

The input variables to the BDT are $\pt^\PGm$, $\pt^{\tauh}$, \mcol, \ptvecmiss, \mttmet, \detamtauh, \dphimtauh, and \dphitauhmet. The neutrino is assumed to be collinear with \tauh, which motivates using the \dphitauhmet variable. The two leptons are usually produced in opposite directions of the azimuthal plane, which motivates using the \dphimtauh variable. The post-fit distributions of simulated signal, data, and backgrounds in each category of the \muhad channel are shown in Fig.~\ref{fig:bdt_muhad}.

\subsection{\texorpdfstring{\ensuremath{\PH \to \mu \taue}\xspace}{Hmutaue}}
In this channel, the preselection requires a muon and electron of opposite electric charge with a separation of $\dr>0.3$. The triggers require both a muon and an electron, where the muon has \pt above 23\GeV, and the electron has \pt above 12\GeV. The muon is required to have $\pt>24\GeV$, $\aeta<2.4$, and $\irelm<0.15$. The electron is required to have $\pt>13\GeV$, $\aeta<2.5$, and $\irele<0.1$. The selections for the \mue channel are summarized in Table~\ref{tab:evtselection_mutau}.

\begin{figure*}[htbp]
  \begin{center}
  \includegraphics[width=0.45\textwidth]{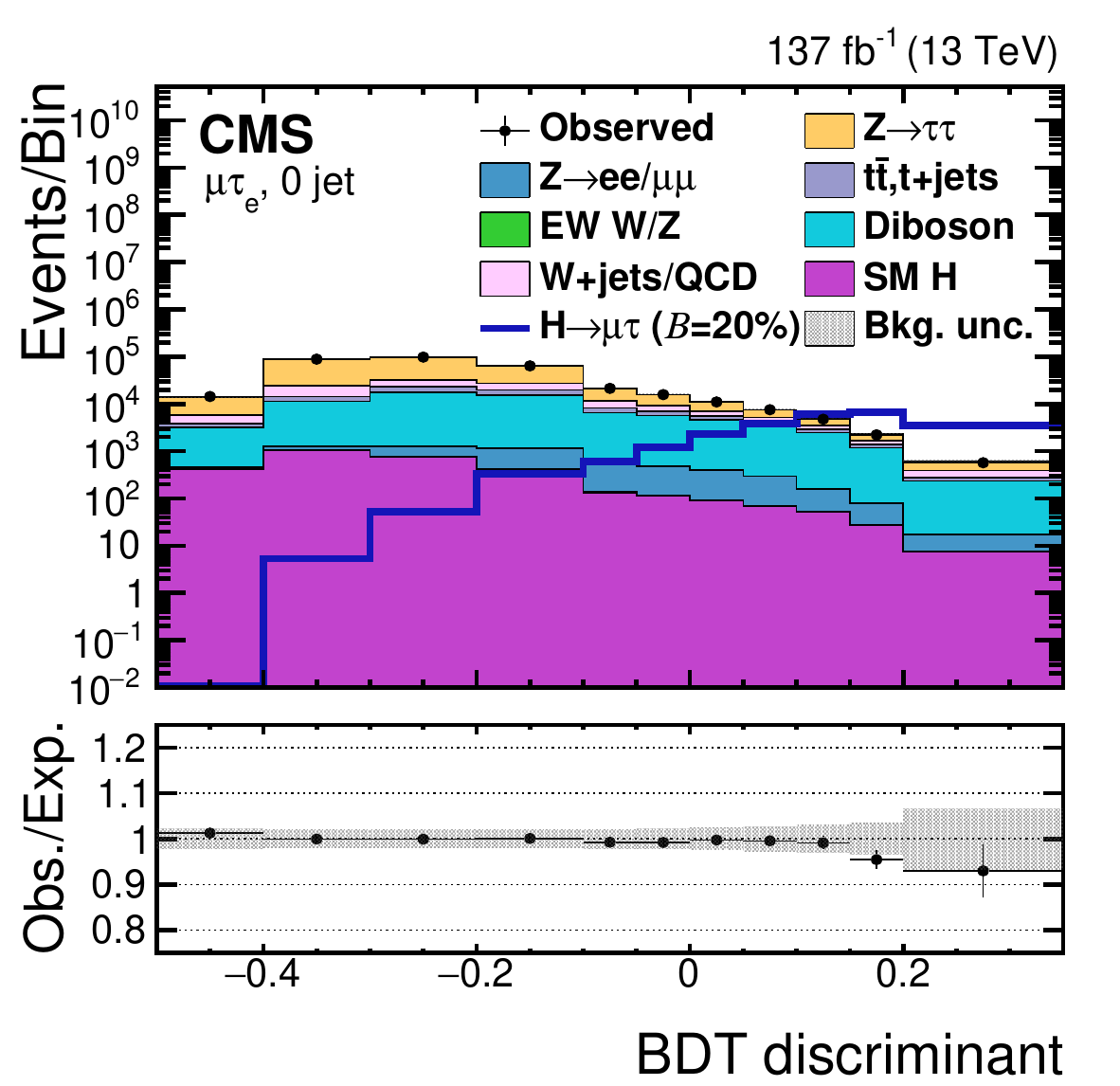}
  \includegraphics[width=0.45\textwidth]{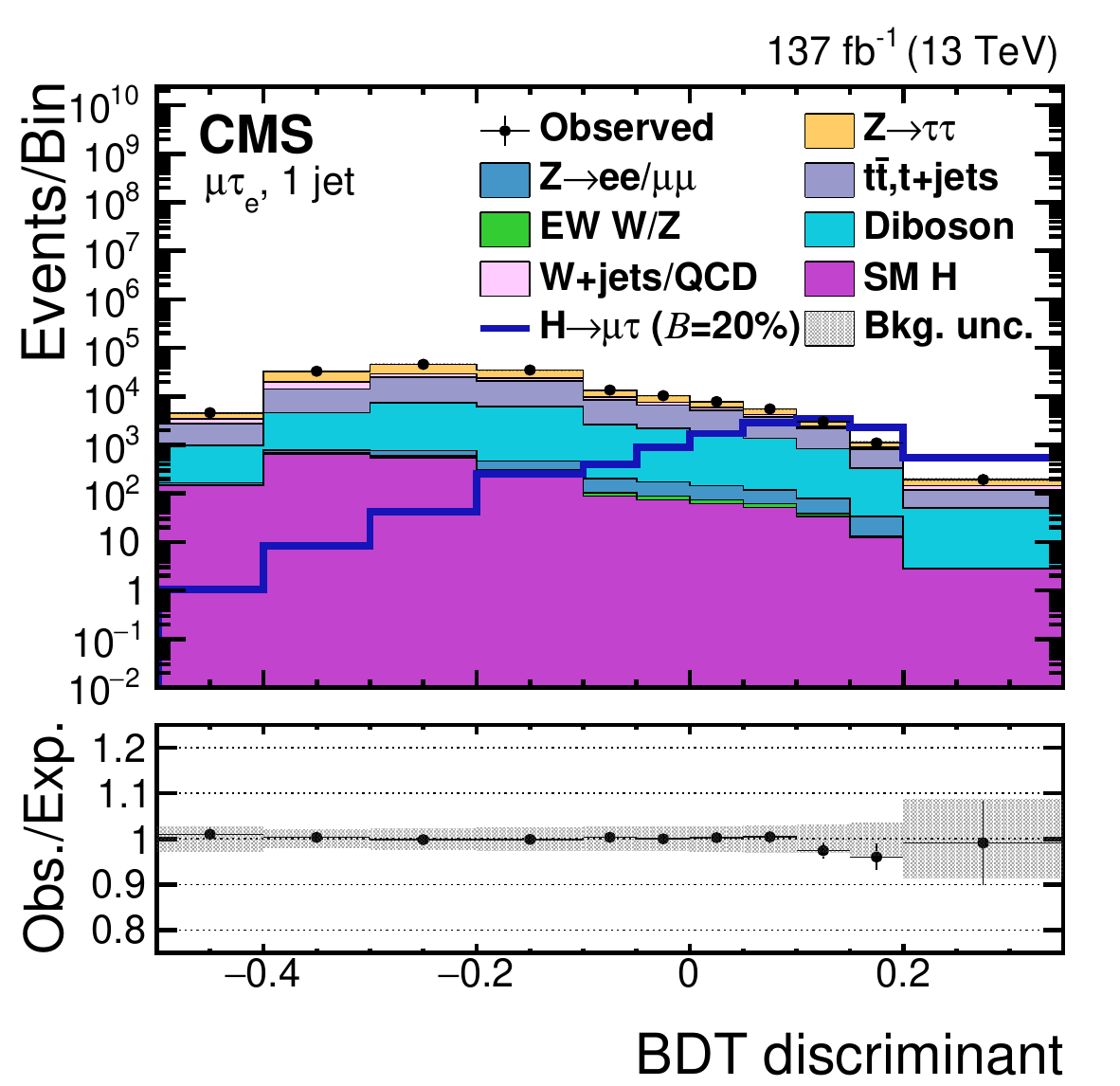}\\
  \includegraphics[width=0.45\textwidth]{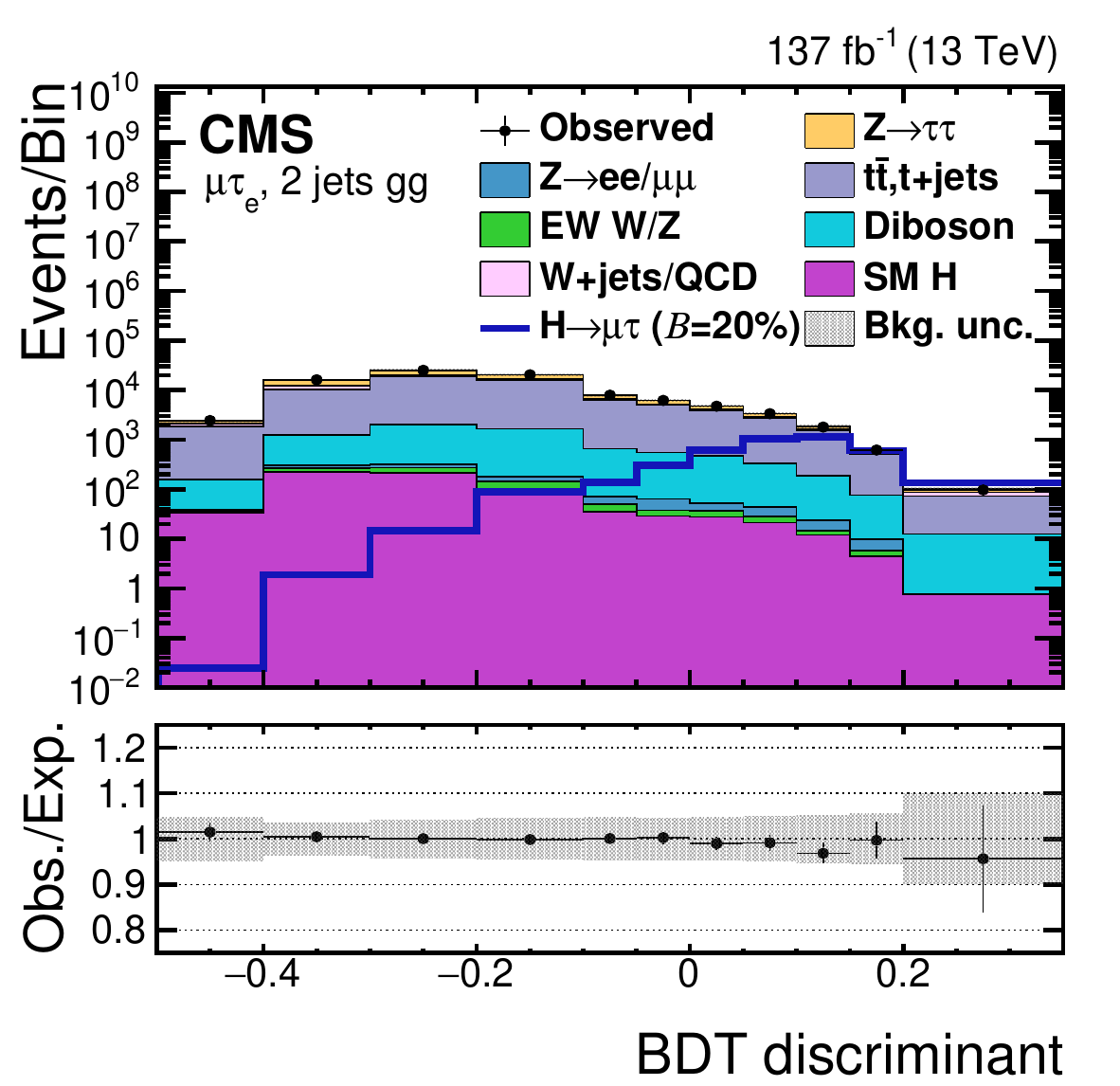}
  \includegraphics[width=0.45\textwidth]{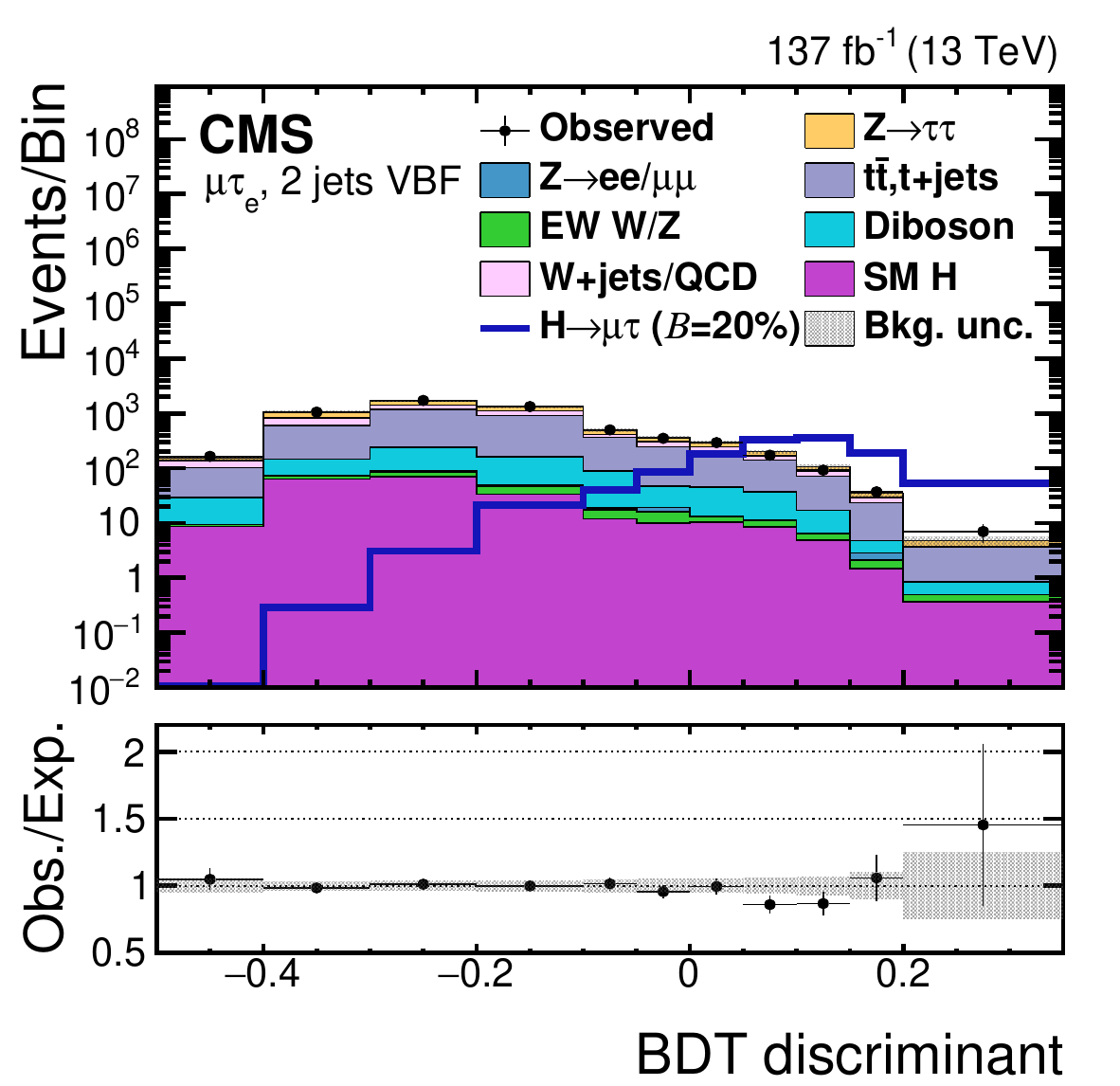}\\
  \caption{BDT discriminant distributions for the data and background processes in the \Hmue channel. A $\BHmt=20\%$ is assumed for the signal. The channel categories are 0 jets (upper row left), 1 jet (upper row right), 2 jets \Pg{}\Pg{}\PH (lower row left), and 2 jets VBF (lower row right). The lower panel in each plot shows the ratio of data and estimated background. The uncertainty band corresponds to the background uncertainty in which the post-fit statistical and systematic uncertainties are added in quadrature.}
  \label{fig:bdt_mue}
  \end{center}
\end{figure*}

The input variables to the BDT are $\pt^\PGm$, $\pt^\Pe$, \mcol, \mtmmet, \mtemet, \dphiem, \dphimmet, and \dphiemet. The neutrinos are assumed to be collinear with the electron, which motivates using the \dphiemet variable. The two leptons are usually produced in opposite directions of the azimuthal plane, which motivates using the \dphiem variable. The post-fit distributions of simulated signal, data, and backgrounds in each category of the \mue channel are shown in Fig.~\ref{fig:bdt_mue}.

\begin{table}[htbp]
\topcaption{Event selection criteria for the \Hmt channels.}
\begin{center}
\begin{scotch}{ccccc}
Variable                        &   \muhad             &  \mue                              \\[0.5ex]
\hline \\ [-2.0ex]
$\pt^{\Pe}$                     &   \NA                &  $>$13\GeV                         \\[0.5ex]
$\pt^{\PGm}$                    &   $>$26\GeV          &  $>$24\GeV                         \\[0.5ex]
$\pt^{\tauh}$                   &   $>$30\GeV          &  \NA                               \\[1.0ex]
$\abs{\eta}^{\Pe}$              &   \NA                &  $<$2.5                            \\[0.5ex]
$\abs{\eta}^{\PGm}$             &   $<$2.1             &  $<$2.4                            \\[0.5ex]
$\abs{\eta}^{\tauh}$            &   $<$2.3             &  \NA                               \\[1.0ex]
$I^{\Pe}_{\text{rel}}$          &   \NA                &  $<$0.1                            \\[0.5ex]
$I^{\PGm}_{\text{rel}}$         &   $<$0.15            &  $<$0.15                           \\[1.0ex]
\multirow{2}{*}{\begin{tabular}{c}Trigger \\ requirement \end{tabular}}
                                & $\pt^{\PGm}>24\GeV$ (all years) &
\multirow{2}{*}{\begin{tabular}{c}$\pt^{\Pe}>12\GeV$ \\ $\pt^{\PGm}>23\GeV$ \end{tabular}}  \\[0.5ex]
                                &    $\pt^{\PGm}>27\GeV$ (2017)   &                         \\[0.5ex]
\end{scotch}
\label{tab:evtselection_mutau}
\end{center}
\end{table}

\subsection{\texorpdfstring{\ensuremath{\PH \to \Pe \tauh}\xspace}{Hetauh}}
In this channel, the preselection requires an electron and \tauh of opposite electric charge with a separation of $\dr>0.5$. The triggers require the presence of an isolated electron with a \pt threshold of 25\GeV (2016), 27\GeV (2017), or 32\GeV (2018). In 2017 and 2018, the signal acceptance is increased by selecting events where the electron has \pt above 24\GeV and the \tauh has \pt above 30\GeV. The electron is required to have $\pt>27\GeV$, $\aeta<2.1$, and $\irele<0.15$. The \tauh is required to have $\pt>30\GeV$ and $\aeta<2.3$. The selections for the \ehad channel are summarized in Table~\ref{tab:evtselection_etau}.

\begin{figure*}[htbp]
  \begin{center}
  \includegraphics[width=0.45\textwidth]{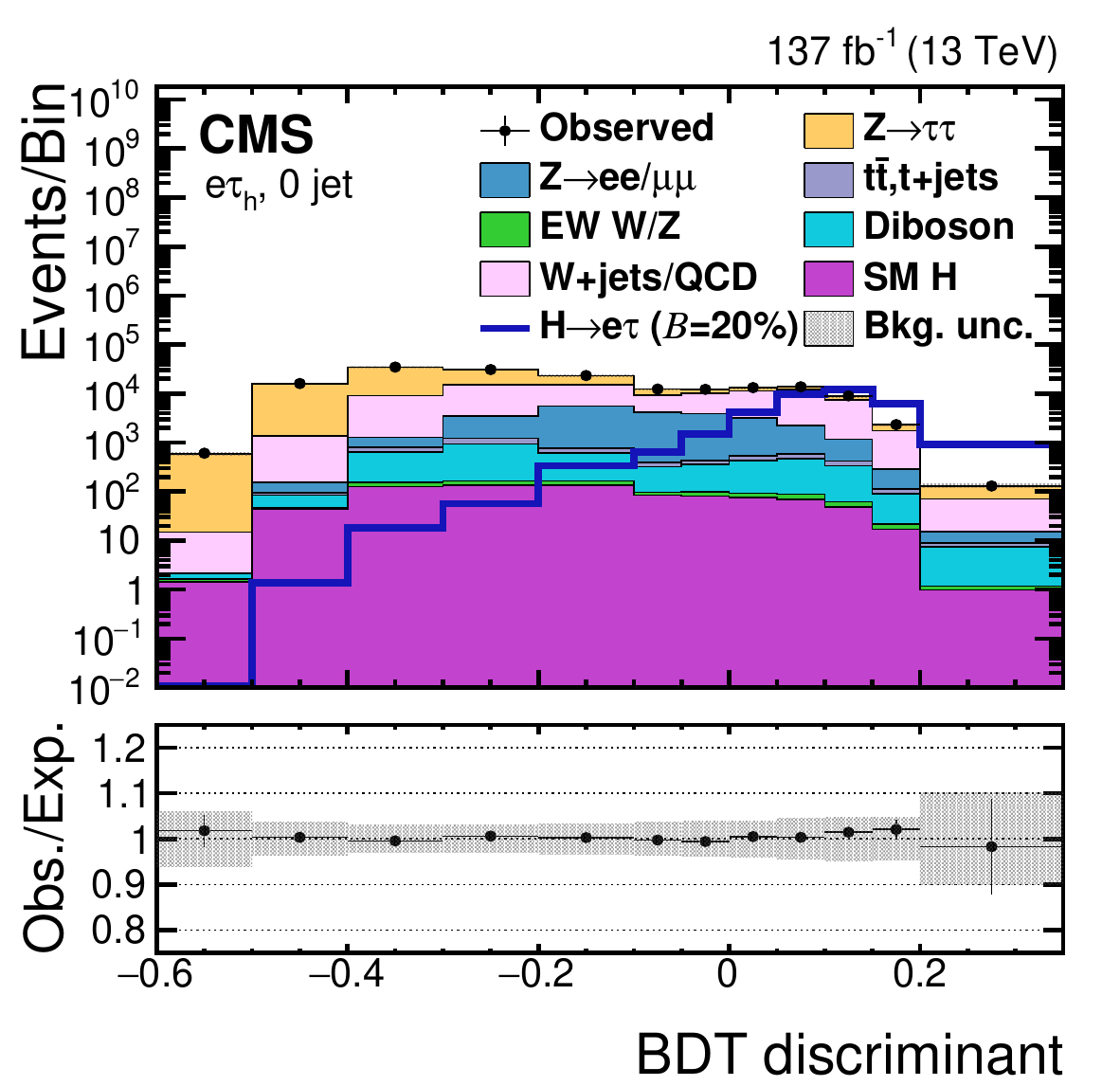}
  \includegraphics[width=0.45\textwidth]{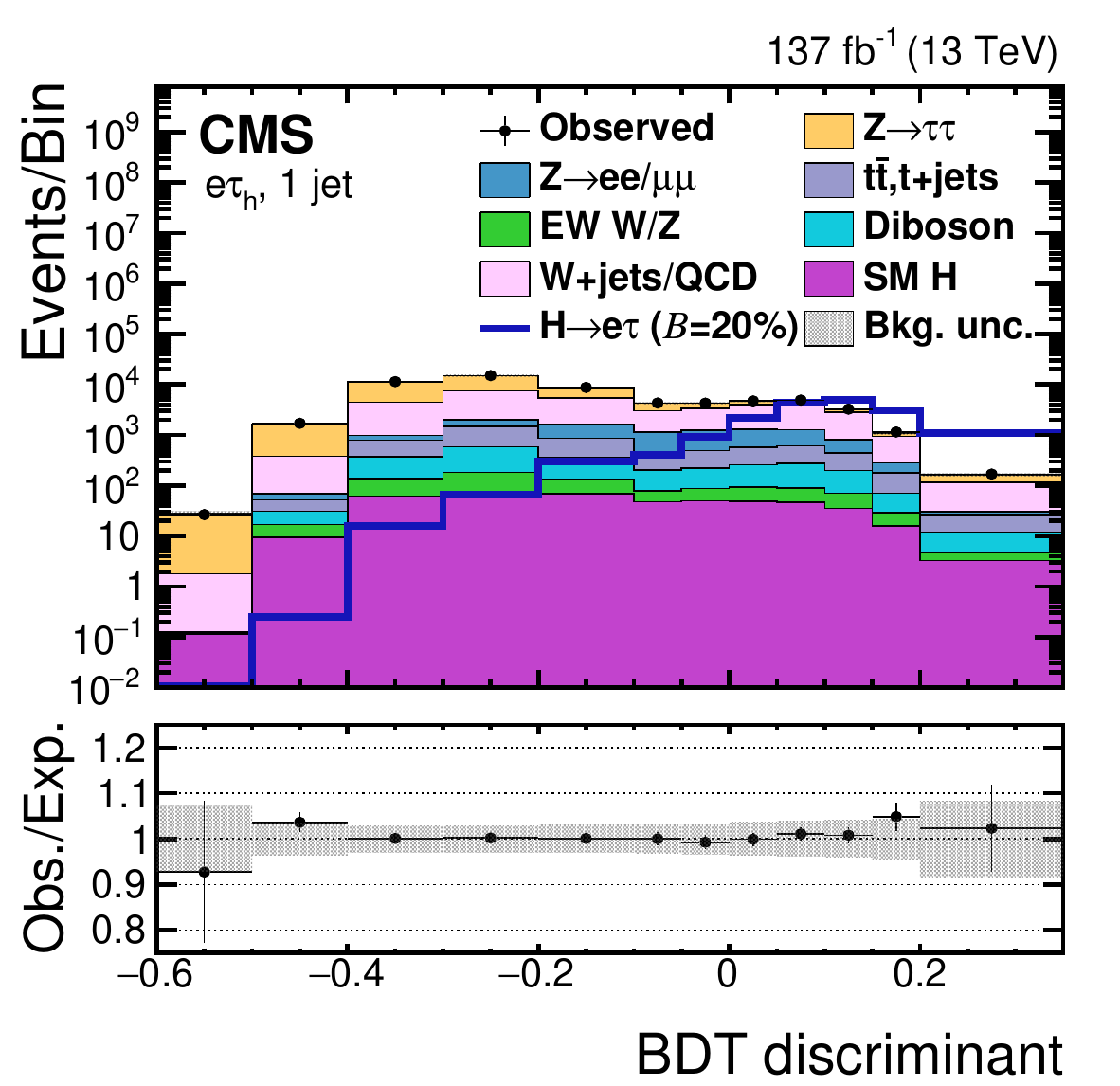}\\
  \includegraphics[width=0.45\textwidth]{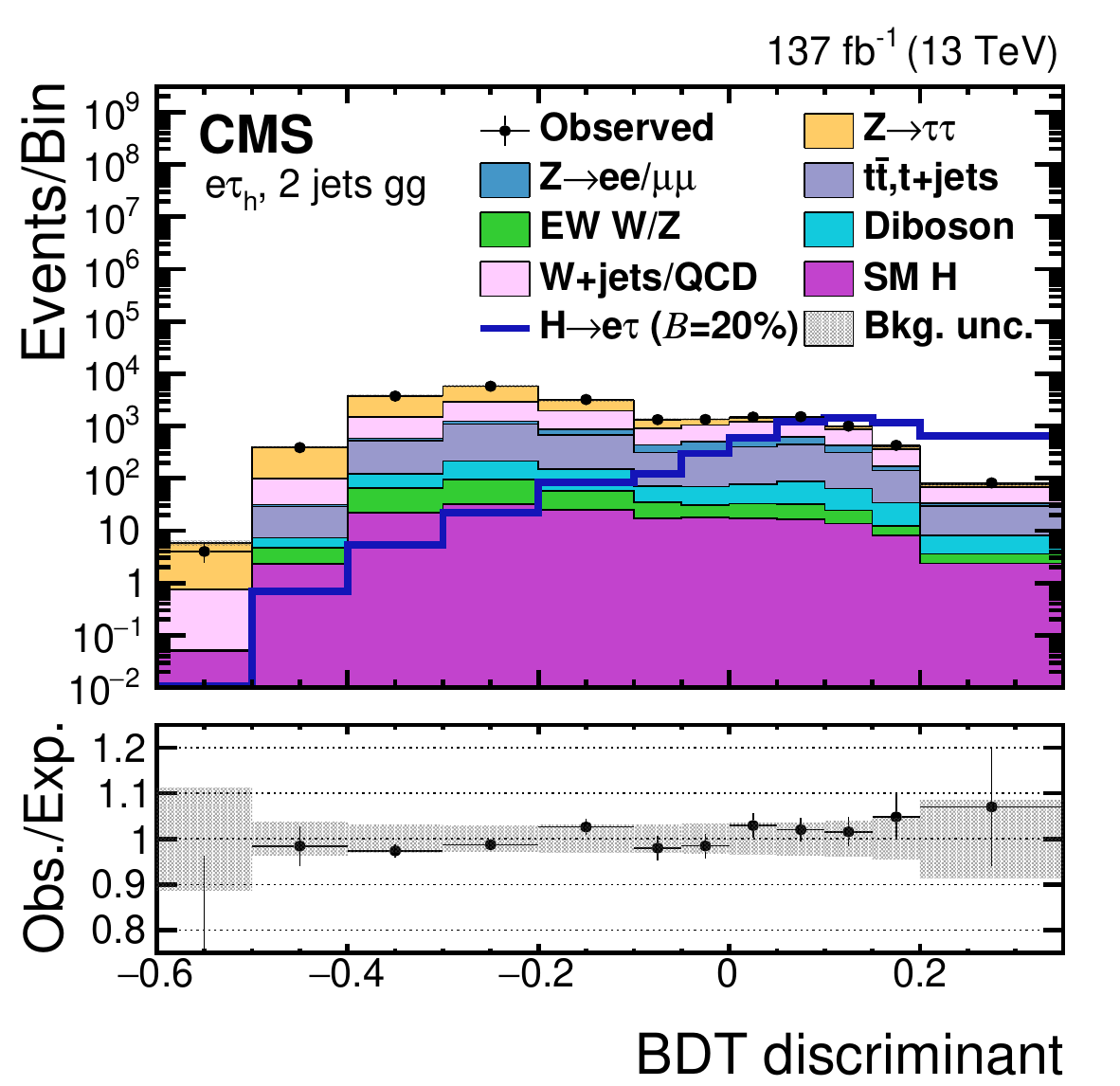}
  \includegraphics[width=0.45\textwidth]{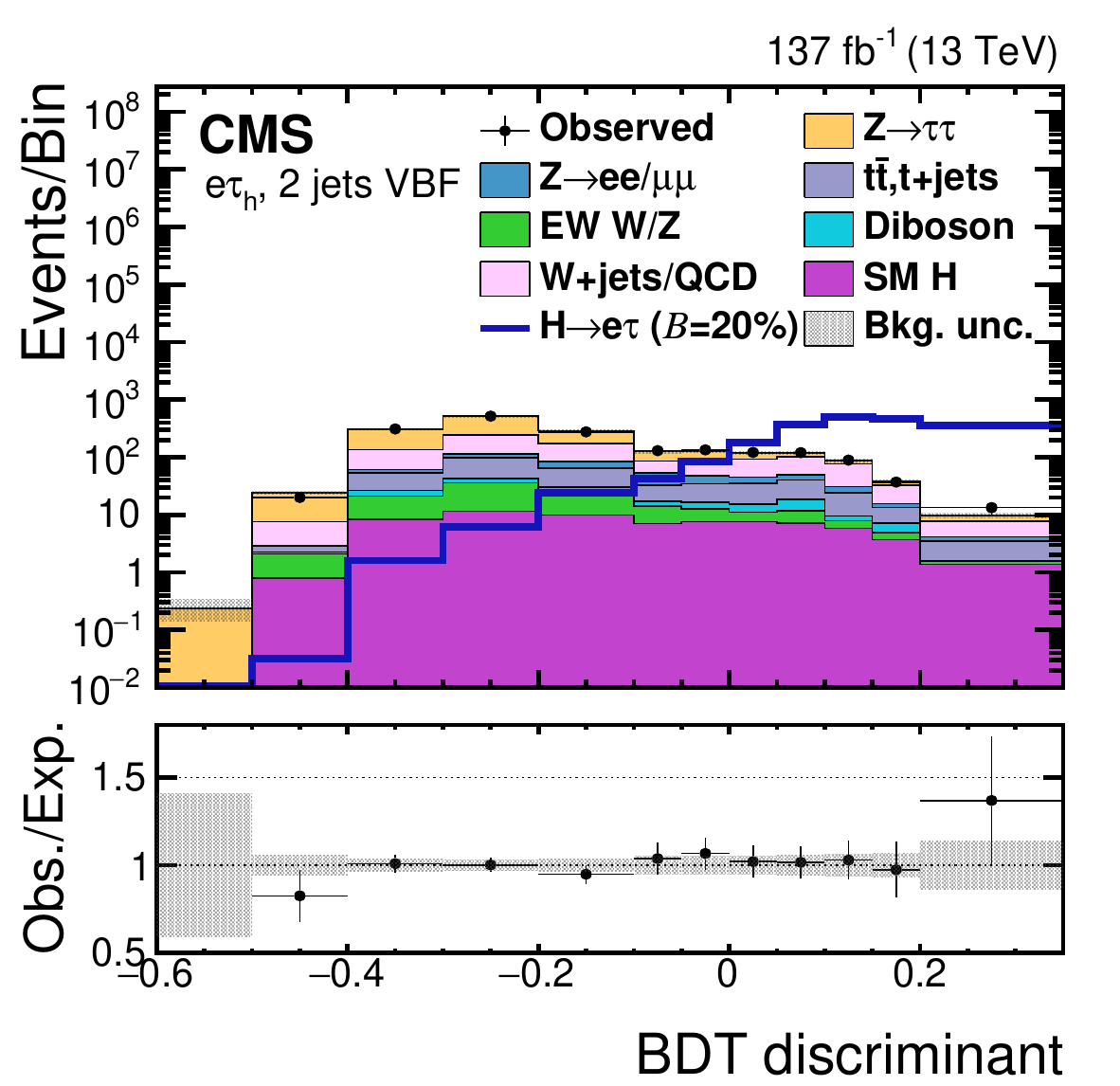}\\
  \caption{BDT discriminant distributions for the data and background processes in the \Hehad channel. A $\BHet=20\%$ is assumed for the signal. The channel categories are 0 jets (upper row left), 1 jet (upper row right), 2 jets \Pg{}\Pg{}\PH (lower row left), and 2 jets VBF (lower row right). The lower panel in each plot shows the ratio of data and estimated background. The uncertainty band corresponds to the background uncertainty in which the post-fit statistical and systematic uncertainties are added in quadrature.}
  \label{fig:bdt_ehad}
  \end{center}
\end{figure*}

The input variables to the BDT are $\pt^\Pe$, $\pt^{\tauh}$, \mcol, \mvis, \mttmet, \detaetauh, \dphietauh, and \dphitauhmet. As can be seen, the input variables are similar to \muhad channel except for the addition of the variable \mvis and removing \ptvecmiss. The variable \mvis has better separation power as the \ehad channel has more $\Zee{+}\text{jets}$ background than the $\Zmm{+}\text{jets}$ background in the \muhad channel. The post-fit distributions of simulated signal, data, and backgrounds in each category of the \ehad channel are shown in Fig.~\ref{fig:bdt_ehad}.

\subsection{\texorpdfstring{\ensuremath{\PH \to \Pe \taum}\xspace}{Hetaumu}}
In this channel, the preselection requires an electron and muon of opposite electric charge with a separation of $\dr>0.4$. The triggers require both an electron and a muon, where the electron has \pt above 23\GeV, and the muon has \pt above 8\GeV. The electron is required to have $\pt>24\GeV$, $\aeta<2.5$, and $\irele<0.1$. The muon is required to have $\pt>10\GeV$, $\aeta<2.4$, and $\irelm<0.15$. The selections for the \emu channel are summarized in Table~\ref{tab:evtselection_etau}.

\begin{figure*}[htbp]
  \begin{center}
  \includegraphics[width=0.45\textwidth]{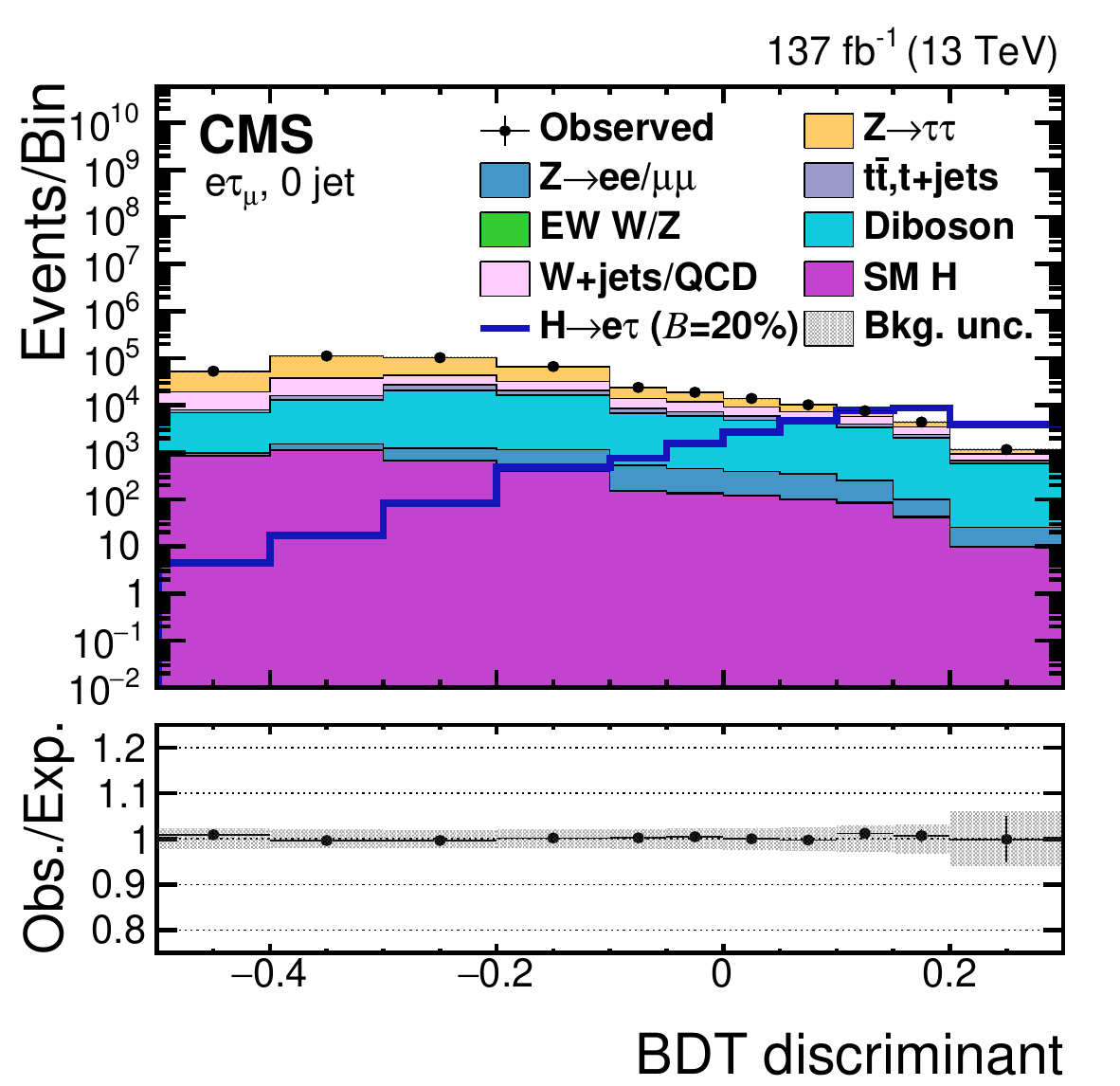}
  \includegraphics[width=0.45\textwidth]{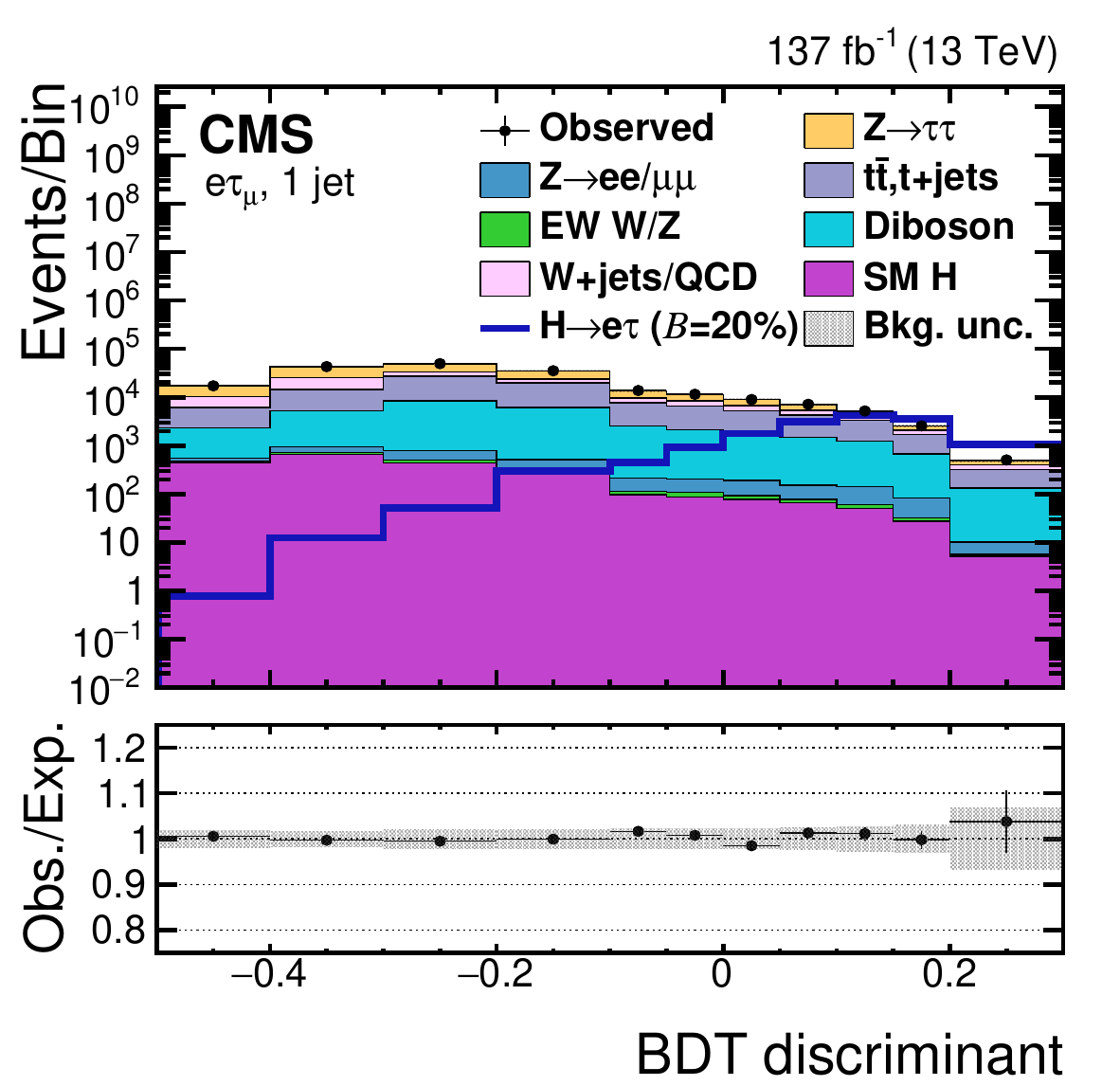}\\
  \includegraphics[width=0.45\textwidth]{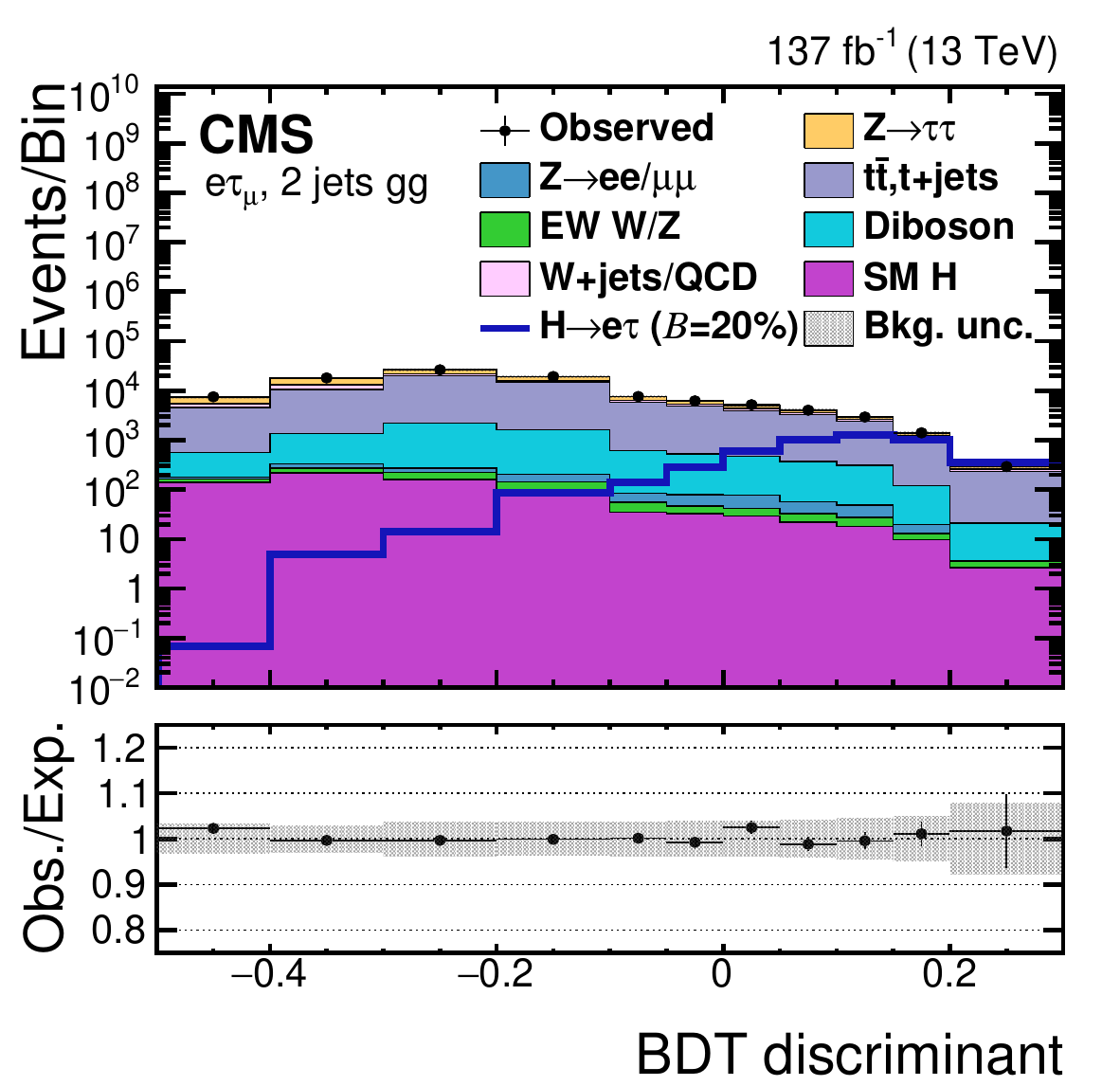}
  \includegraphics[width=0.45\textwidth]{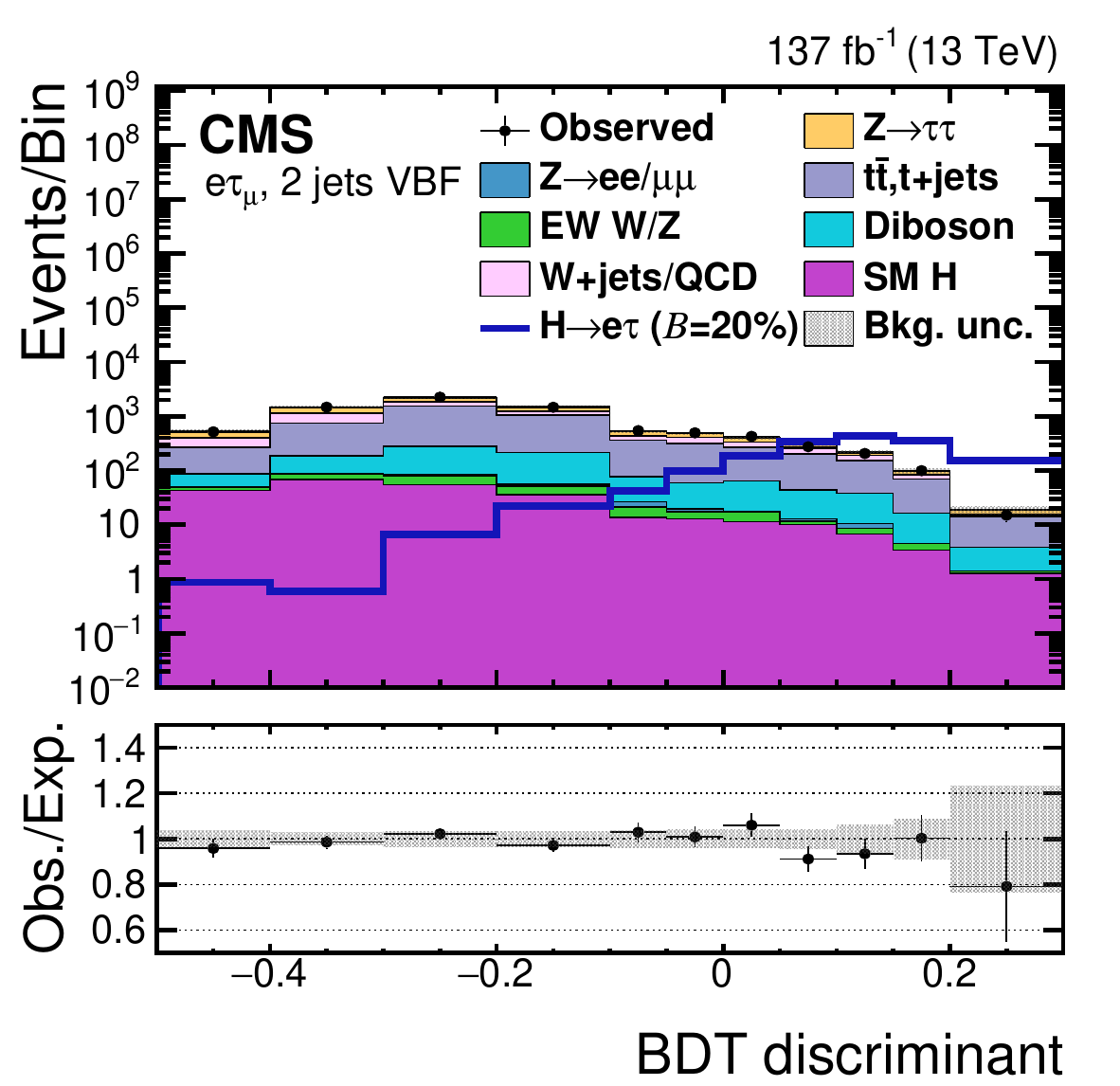}\\
  \caption{BDT discriminant distributions for the data and background processes in the \Hemu channel. A $\BHet=20\%$ is assumed for the signal. The channel categories are 0 jets (upper row left), 1 jet (upper row right), 2 jets \Pg{}\Pg{}\PH (lower row left), and 2 jets VBF (lower row right). The lower panel in each plot shows the ratio of data and estimated background. The uncertainty band corresponds to the background uncertainty in which the post-fit statistical and systematic uncertainties are added in quadrature.}
  \label{fig:bdt_emu}
  \end{center}
\end{figure*}

The input variables to the BDT are $\pt^\PGm$, $\pt^\Pe$, \mcol, \mvis, \mtmmet, \dphiem, \dphimmet, and \dphiemet. As can be seen, the input variables are similar to \mue channel except for the addition of the variable \mvis and removing \mtemet. The post-fit distributions of simulated signal, data, and backgrounds in each category of the \emu channel are shown in Fig.~\ref{fig:bdt_emu}.

\begin{table*}[htbp]
\topcaption{Event selection criteria for the \Het channels.}
\begin{center}
\begin{scotch}{ccccc}
Variable                        &          \ehad                     &  \emu                           \\[0.5ex]
\hline \\ [-2.0ex]
$\pt^{\Pe}$                     &          $>$27\GeV                 &  $>$24\GeV                      \\[0.5ex]
$\pt^{\PGm}$                    &          \NA                       &  $>$10\GeV                      \\[0.5ex]
$\pt^{\tauh}$                   &          $>$30\GeV                 &  \NA                            \\[1.0ex]
$\abs{\eta}^{\Pe}$              &          $<$2.1                    &  $<$2.5                         \\[0.5ex]
$\abs{\eta}^{\PGm}$             &          \NA                       &  $<$2.4                         \\[0.5ex]
$\abs{\eta}^{\tauh}$            &          $<$2.3                    &  \NA                            \\[1.0ex]
$I^{\Pe}_{\text{rel}}$          &          $<$0.15                   &  $<$0.1                         \\[0.5ex]
$I^{\PGm}_{\text{rel}}$         &          \NA                       &  $<$0.15                        \\[1.0ex]
\multirow{4}{*}{\begin{tabular}{c}Trigger \\ requirement \end{tabular}}
                                &   $\pt^{\Pe}>25\GeV$ (2016)        &
\multirow{4}{*}{\begin{tabular}{c}$\pt^{\Pe}>23\GeV$ \\ $\pt^{\PGm}>8\GeV$ \end{tabular}}              \\[0.5ex]
                                &   $\pt^{\Pe}>27\GeV$ (2017)        &                                 \\[0.5ex]
                                &   $\pt^{\Pe}>32\GeV$ (2018)        &                                 \\[0.5ex]
                                & $\pt^{\Pe}>24\GeV$ and $\pt^{\tauh}>30\GeV$ (2017, 2018) &           \\[0.5ex]
\end{scotch}
\label{tab:evtselection_etau}
\end{center}
\end{table*}

\section{Background estimation}
\label{sec:bkgEstimation}

One of the major background contributions comes from the \Ztt process, in which the muon or electron arises from a \PGt lepton decay. The other major background contributions arise from the \wjets process and from multijets events produced through the strong interaction (referred to as QCD multijet events hereafter), where one or more of the jets are misidentified as leptons. These backgrounds are estimated from data either fully or with the aid of simulation. The \ttbar and single top quark background contributes substantially in leptonic channels and is estimated using simulated events along with the other backgrounds. The background estimates are validated in different orthogonal VRs constructed to have enhanced contributions from specific backgrounds.

\subsection{\texorpdfstring{\ensuremath{\PZ \to \PGt \PGt}\xspace}{Ztt} background}
The \Ztt background is estimated from data using an embedding technique~\cite{Sirunyan:2019drn}. This technique allows for an estimation of the genuine \PGt{}\PGt SM backgrounds from data with reduced simulation input. This minimizes the uncertainties that arise from using simulation. Events with a pair of oppositely charged muons are selected in data so that \Zmm events largely dominate. These data events are selected independently of the event selection criteria described in Section~\ref{sec:selection}. The muons are removed from the selected events and replaced with simulated \PGt leptons with the same kinematic properties as those of the replaced muon. In that way, a set of hybrid events is obtained that relies on simulation only for the decay of the \PGt leptons. The description of the underlying event or the production of associated jets is taken entirely from data. This technique results in a more accurate description of the \ptvecmiss and jet-related variables than simulation and an overall reduction in the systematic uncertainties. Embedded events cover all backgrounds with two genuine \PGt leptons, and this includes a small fraction of \ttbar, diboson, and EW \PW/\PZ events. The simulated events from the \ttbar, diboson, and EW \PW/\PZ where both \PGt candidates match to \PGt leptons at the generator level are removed to avoid any double counting.

\subsection{Misidentified lepton background}
The misidentified lepton background corresponds to events where jets are misidentified as leptons. They mostly arise from two sources: \wjets and QCD multijet events. In \wjets background events, one of the leptons is from the \PW boson decay while the other is a jet misidentified as a lepton. In QCD multijet events, both the leptons are misidentified jets. In the \muhad and \ehad channels, the contributions from misidentified lepton backgrounds have been estimated using a ``misidentification rate'' approach. In the \mue and \emu channels, an ``extrapolation factor'' approach is adopted, which is consistent with the ``misidentification rate'' approach, and is used because of limited statistical precision in the leptonic channels.

\subsubsection{Misidentification rate approach}
The misidentified lepton background in the signal region (SR) is estimated using misidentification rates from \zjets CR and applied to a background-enriched region from collision data. The misidentification rates are evaluated using events with a \PZ boson and at least one jet that can be misidentified as a lepton. The probabilities with which jets are misidentified as an electron, muon, or \tauh are labeled as $f_\Pe, f_\PGm$, and $f_{\tauh}$, respectively. The \PZ boson is formed using two muons with $\pt>26\GeV$, $\aeta<2.4$, and $\irelm<0.15$ for measuring the jet $\to\tauh,\PGm,\Pe$ misidentification rate. The muons are required to be oppositely charged and have invariant mass between 70 and 110\GeV. The contribution from diboson events, where there is a genuine lepton, is subtracted using simulation.

The jet is required to pass the same lepton identification criteria as used in the SR. A ``signal-like'' and ``background-like'' regions are defined. The isolation for the electron and muon is required to have $\irel<0.15$ and the \tauh discriminated against jets at a WP that has an identification efficiency of about 70\% for the ``signal-like'' region. For the ``background-like'' region, lepton isolation is required to be $0.15<\irelm<0.25$ and $0.15<\irele<0.50$, and the \tauh is discriminated against jets at a WP that has an identification efficiency of about 80\% and not pass the WP that has an identification efficiency of about 70\%. After the ``signal-like'' and ``background-like'' regions are defined, the misidentification rates are computed as functions of the lepton \pt. The misidentification rates $f_\Pe, f_\PGm,$ and $f_{\tauh}$ are estimated as:
\begin{linenomath*}
  \begin{equation*}
    f_i = \frac{S_i}{B_i + S_i}
  \end{equation*}
\end{linenomath*}
where $S_i$ is the number of events in the ``signal-like'' region, while $B_i$ is the number of events in the ``background-like'' region. The \tauh misidentification rate shows a \pt dependence that depends on the \tauh decay mode and \aeta and is therefore evaluated as a function of $\pt^\PGt$ for the different decay modes and two $\eta$ regions ($\aeta<1.5$ or $\aeta>1.5$).

In the \ehad channel, the \tauh misidentification rate is evaluated using events with a \PZ boson formed using two electrons with $\pt>27\GeV$, $\aeta<2.5$, and $\irele<0.15$. The electrons must be oppositely charged and have an invariant mass between 70 and 110\GeV. The reason for using \Zee events for evaluating the \tauh misidentification rate in \ehad channel is that the DNN WPs used for discriminating \tauh against electrons and muons are different in this channel compared to the \muhad channel as described in Section~\ref{sec:event_reco}. The misidentification rates evaluated using this CR are compatible with the misidentification rates measured in \Zmm events.

The computed misidentification rates $f_i$ depend on the lepton \pt for electrons and muons or \pt, $\eta$, and decay mode for the \tauh candidates. The misidentification rates for electrons and muons are ${\approx}0.4$ and ${\approx}0.6$, respectively, at $\pt=30.0\GeV$. The misidentification rates for \tauh candidates are in the range 0.02--0.24 at $\pt=30.0\GeV$. They are used to estimate the background yields and obtain the distributions of the misidentified lepton background. This is accomplished through the following procedure. Each event in the ``background-like'' region, defined using the collision data with the same selection as the SR, but loosening the isolation requirements on one of the leptons, is weighted by a factor $f_i/(1-f_i)$. Events with the possibility of double counting because of two misidentified leptons are subtracted. For example, events with both a misidentified muon or electron and a misidentified \tauh are accounted once in the ``background-like'' region for muon or electron with a weight $f_\ell/(1-f_\ell)$ and another time in the ``background-like'' region for \tauh with a weight $f_\PGt/(1-f_\PGt)$ and are hence double counted. This is mitigated by subtracting their contribution once using a weight, $f_\PGt f_\ell/[(1-f_\PGt)(1-f_\ell)]$, where $\ell=\PGm\,\text{or}\,\Pe$.

The background estimate is validated in a VR by requiring the two leptons to have the same electric charge, enhancing the misidentified lepton background. Figure~\ref{fig:mutau_control} (left) shows the comparison of data with background estimates in this VR for the \muhad channel. The background estimate is also validated in a \PW boson enriched VR, as shown in Fig.~\ref{fig:mutau_control} (middle). This VR is obtained by applying the preselection, $\mT(\ell,\ptvecmiss)>60\GeV$ ($\ell=\Pe$ or \PGm), and $\mT(\tauh,\ptvecmiss)>80\GeV$.

\subsubsection{Extrapolation factor approach}
In the \emu and \mue channels, the QCD multijet background is estimated from the data using events with an electron and a muon with the same electric charge~\cite{Sirunyan:2017khh}. Contributions from other processes are estimated from simulation and subtracted from the data. Extrapolation factors from the CR requiring the two leptons to have the same electric charge to the SR are measured in data as a function of the jet multiplicity and the \dr separation between the electron and muon.

The extrapolation factors are estimated using events with a muon failing the isolation requirement and an isolated electron. The contribution from \bbbar events to the QCD multijet background gives rise to the \dr dependence and is parameterized with a linear function. The extrapolation factors are higher for events with low \dr separation between the electron and muon, decreasing as the \dr separation increases. The extrapolation factors also depend on the electron and muon \pt. This \pt dependence comes from the leptons arising from the semi-leptonic \PQc quark decay. These leptons tend to be softer in \pt and less isolated, resulting in a reduction in the number of such events passing the \pt and isolation requirements.

As the extrapolation factors are from CR where the muon fails the isolation requirement, an additional correction is applied to cover a potential mismodeling. This correction is calculated by measuring the extrapolation factors in two different CRs. The first CR has events where the muon is isolated, and the electron fails the isolation requirement. The second CR has events where both the electron and muon fail the isolation requirement. The ratio of the extrapolation factors from these two CRs is taken as the correction to account for the potential mismodeling induced by requiring the muon to fail the isolation requirement.

\subsection{Other backgrounds}
\begin{figure*}[htbp]
  \begin{center}
  \includegraphics[width=0.32\textwidth]{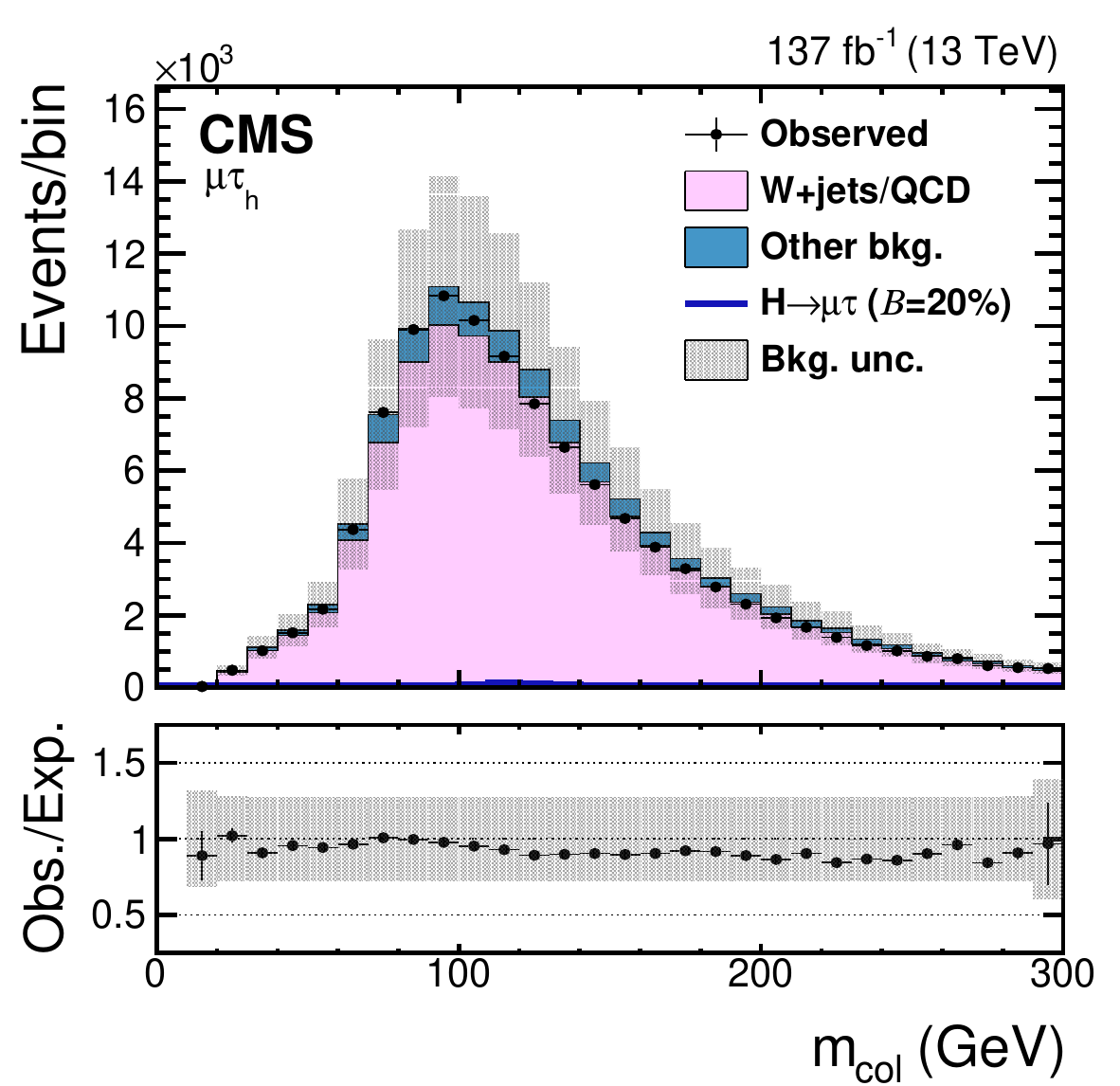}
  \includegraphics[width=0.32\textwidth]{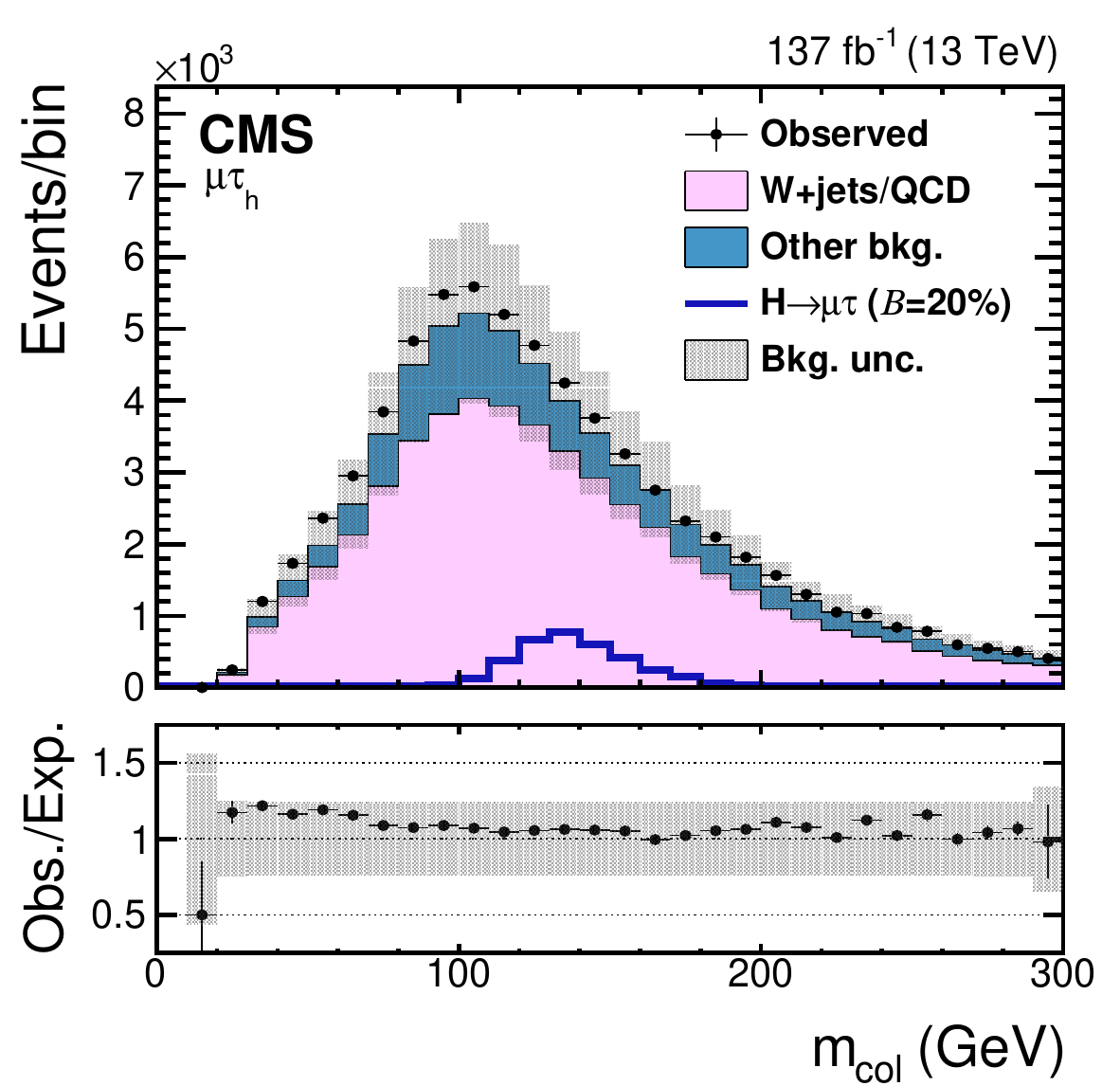}
  \includegraphics[width=0.32\textwidth]{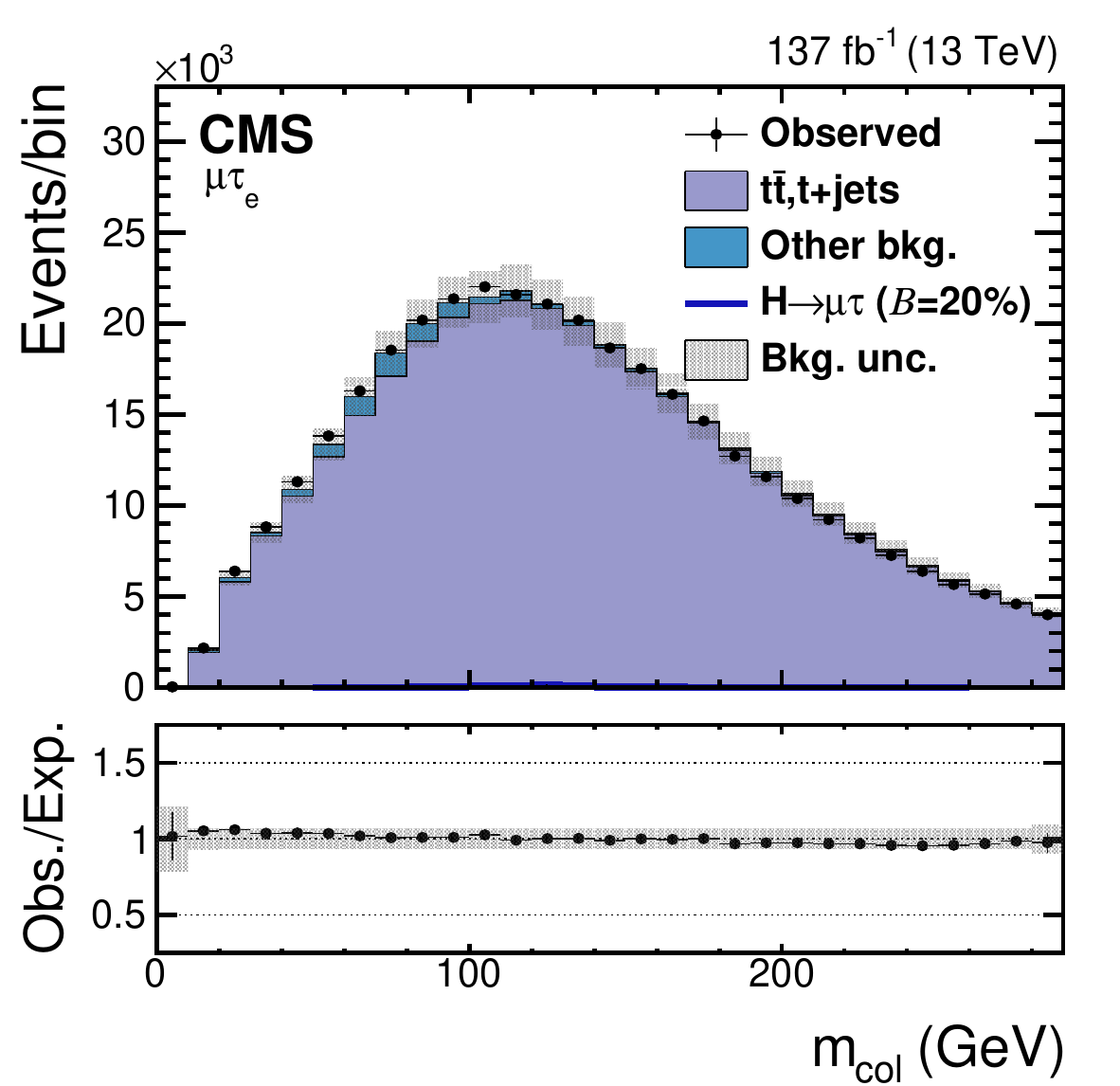}
  \caption{The \mcol distribution in VR with same electric charge for both leptons (left), \wjets VR (middle), and \ttbar VR (right). In each distribution, the VR's dominant background is shown, and all the other backgrounds are grouped into ``Other bkg.''. A $\BHmt=20\%$ is assumed for the signal. The lower panel in each plot shows the ratio of data and estimated background. The uncertainty band corresponds to the background uncertainty in which the post-fit statistical and systematic uncertainties are added in quadrature.}
  \label{fig:mutau_control}
  \end{center}
\end{figure*}

Other background contributions come from processes in which a lepton pair is produced from the weak decays of quarks and vector bosons. These include \ttbar, \PW{}\PW, \PW{}\PZ, and \PZ{}\PZ events. There are nonnegligible contributions from processes such as $\PW\Pgg^{(\ast)}{+}\text{jets}$, single top quark production, and \Zll $(\ell=\Pe,\PGm)$. Figure~\ref{fig:mutau_control} (right) shows the comparison of data with background estimates in the \ttbar VR for the \mue channel. This VR is defined by requiring the presence of at least one {\cPqb}-tagged jet in the event in addition to the preselection. The SM Higgs boson production contribution mainly comes from \Htt and \HWW decays.

\section{Systematic uncertainties}
\label{sec:sysUnc}

Several sources of experimental and theoretical systematic uncertainties are taken into account in the statistical analysis. These uncertainties affect both the normalization and distribution of the different processes. The different systematic uncertainties are incorporated in the likelihood as nuisance parameters for which log-normal a priori distributions are assumed, and distribution variations are taken into account via continuous morphing~\cite{Conway:2011in}. The maximum likelihood and profile likelihood with asymptotic approximation are then computed using the defined likelihood to obtain the best fit branching fraction and upper limits on the branching fraction for the LFV Higgs boson decays. As the search is categorized into different final states, partial and complete correlations between the uncertainties in different categories are taken into account and are summarized in Table~\ref{tab:systematics}.

The uncertainties to reconstruct a \tauh and estimate its identification efficiency for different \pt ranges are measured using a tag-and-probe method~\cite{Khachatryan:2010xn} and found to be in the range of 2--3\%. The uncertainties for different ranges of \pt are treated as uncorrelated. These uncertainties are also considered for the embedded \PGt{}\PGt background, where they are treated as 50\% correlated with the simulation uncertainties. For the embedded events, triggering on muons before being replaced by \PGt leptons leads to an uncertainty in the trigger efficiency of about 4\%, which is treated as uncorrelated between the three years due to different triggering criteria. There are two effects that need to be considered for the embedded events. The embedded events have higher track reconstruction efficiency because of reconstruction in an empty detector environment. The energy deposits of the replaced muons can cause event migration for \tauh decay modes with a $\pi^{0}$. Data to simulation scale factors cover these effects with corresponding systematic uncertainties.

Uncertainties arising from an electron or a muon misidentified as \tauh correspond to between 7--40\% or 10--70\%, respectively, for different bins of \pt, $\eta$, and \tauh decay modes. The uncertainty in the \tauh energy scale is treated as uncorrelated for different decay modes and 50\% correlated between embedded and simulated backgrounds and ranges from 0.7--1.2\%. The uncertainty in the electron energy scale and the muon momentum scale for misidentified leptons is independent of the \tauh energy scale and amounts to 7\% and 1\%, respectively. The effect of lepton energy resolution is found to be negligible.

The jet energy scale is affected by several sources, and its uncertainty is evaluated as a function of \pt and $\eta$. The jet energy scale's effect is propagated to the BDT discriminant and varies from 3--20\%~\cite{Khachatryan:2016kdb}. The uncertainties in jet energy resolution are also taken into account and mostly impact the \mjj-defined categories. The jets with $\pt<10\GeV$ fall under unclustered energy. The unclustered energy scale is considered independently for charged particles, neutral hadrons, photons, and very forward particles, that affect both the distributions and the total yields and are treated as uncorrelated. The efficiency to classify a jet as {\cPqb}-tagged is different in data and simulation, and scale factors that depend on jet \pt are used to correct the simulation. The uncertainties in the measured values of these scale factors are taken as sources of systematic uncertainties.

The uncertainties in the reconstruction of electrons and muons, along with their isolation criteria, are measured using the tag-and-probe method in data in \Zee and \Zmm events and sum up to about 2\%~\cite{Chatrchyan:2012xi, Khachatryan:2015hwa, Khachatryan:2015dfa}. The uncertainty in the measurement of the muon momentum scale is in the range 0.4--2.7\% for different \aeta ranges, while for the electron momentum scale, it is less than 1\%. The selection of events using electron- and muon-based triggers results in an additional 2\% uncertainty in the yield of simulated processes. In the \ehad channel, an additional 5\% uncertainty is associated with using the trigger requiring the presence of both an electron and \tauh in 2017 and 2018. The uncertainties related to the lepton identification and momentum scale are treated as correlated between the three years, while the uncertainties related to the triggering are treated as uncorrelated.

The misidentification rates in the \ehad and \muhad final states are parameterized using a linear function dependent on \tauh \pt, where two uncertainties are ascribed per fit function. The normalization uncertainties in the estimates of the misidentified lepton backgrounds ($\text{jet}\to\tauh,\PGm,\Pe$) from data are taken from the VR, which is defined orthogonally to the SR. Additional uncertainty is estimated for the misidentified lepton background in the \PW boson enriched VR. It is parameterized as a function of \dphimmet for the \muhad channel and as a function of \dphiemet for the \ehad channel. Discriminants with different signal-to-background ratios are used to differentiate \tauh against electrons and muons, which entails an additional 3\% uncertainty for the \ehad channel.

The misidentified lepton background in the \emu and \mue final states is affected by different uncertainties. The statistical uncertainties arising from both fits of the extrapolation factors as a function of the lepton \pt and the spatial separation between electron and muon are taken into account. The uncertainty in extrapolation factors resulting from inverting the muon isolation is taken into account. These uncertainties have a combined effect of about 20\% on the normalization. The dominant source of uncertainty in the simulated background processes, \Zee, \Zmm, \Ztt, \PW{}\PW, \PZ{}\PZ, $\PW\Pgg$, \ttbar, and single top quark production is the measurement of the cross section for these processes and is treated as correlated between the three years.

The theoretical uncertainties affecting the measurement of Higgs boson production cross section are the QCD scales (renormalization and factorization scales), the choice of PDFs, and the strong coupling constant (\alpS) evaluated at the \PZ boson mass. These uncertainties affect the signal's normalization and are treated as correlated between the three years~\cite{deFlorian:2016spz}. The changes made in QCD scales provide 3.9, 0.5, 0.9, and 0.8\% uncertainties in the \Pg{}\Pg{}\PH, VBF, \PZ{}\PH, \PW{}\PH cross sections, respectively, while changes in the PDFs and \alpS result in 3.2, 2.1, 1.3, and 1.9\% uncertainties, respectively. The acceptance is taken into account when changes are made in QCD scales and the PDFs and \alpS.

The normalization of the event yield for \Htt is taken from simulation. The uncertainty in the \BHtt includes a 1.70\% due to missing higher-order corrections, a 0.99\% in the quark masses, and a 0.62\% on \alpS. The normalization of the event yield for \HWW is taken from simulation. The uncertainty in the \BHWW includes a 0.99\% due to missing higher-order corrections, a 0.99\% in the quark masses, and a 0.66\% in \alpS.

The bin-by-bin uncertainties account for the statistical uncertainties in each bin of the distributions of every process. The Barlow--Beeston Lite~\cite{Barlow:1993dm} approach is used, assigning a single parameter to scale the sum of the process yields in each bin, constrained by the total uncertainty, instead of requiring separate parameters, one per process. This is useful to reduce the number of parameters required in the statistical analysis. They are treated as uncorrelated between bins, categories, and channels.

The integrated luminosities of the 2016, 2017, and 2018 data taking periods are individually known to have uncertainties in the 2.3--2.5\% range~\cite{CMS-PAS-LUM-17-001, CMS-PAS-LUM-17-004, CMS-PAS-LUM-18-002}, while the total integrated luminosity has an uncertainty of 1.8\%, the improvement in precision reflecting the uncorrelated time evolution of some systematic effects. The uncertainty in the integrated luminosity affects all processes, with the normalization taken directly from the simulation. Uncertainty related to pileup is evaluated through changes made in the weights applied to the simulation and is treated as correlated between the three years. The dependence on weight is obtained through a 5\% change in the total inelastic cross section used to estimate the number of pileup events in data. Other minimum-bias event modeling and initial- and final-state radiation uncertainties are estimated to be much smaller than those on the rate and are therefore neglected.

During the 2016 and 2017 data taking periods, a gradual shift in the timing of the inputs from the ECAL first-level trigger in the region of $\aeta>2.0$ caused a specific trigger inefficiency. For events containing an electron or a jet with respective \pt $>50\GeV$ or $>100\GeV$, in the region $2.5<\aeta<3.0$ the efficiency loss is 10--20\%, depending on \pt, $\eta$, and time. Correction factors are computed from data and applied to the acceptance evaluated through simulation. Uncertainty due to this correction factor is ${\approx}1\%$ and is treated as correlated between the two years.

\begin{table*}[htbp]
\centering
\topcaption{Systematic uncertainties in the expected event yields. All uncertainties are treated as correlated among categories, except those with two values separated by the $\oplus$ sign. In this case, the first value is the correlated uncertainty and the second value is the uncorrelated uncertainty for each category.}
\cmsTable{
\begin{scotch}{l*{4}{c}}
Systematic uncertainty                   &        \muhad         &        \mue        &        \ehad            &        \emu        \\
\hline
Muon ident. and iso.                     &          2\%          &        2\%         &         \NA             &        2\%         \\
Electron ident. and iso.                 &         \NA           &        2\%         &         2\%             &        2\%         \\
Trigger                                  &          2\%          &        2\%         &         2\%             &        2\%         \\
\tauh ident.                             &   \pt dep. (2--3\%)   &        \NA         &   \pt dep. (2--15\%)    &        \NA         \\
$\PGm\to\tauh$ misid.                    &       10--70\%        &        \NA         &         \NA             &        \NA         \\
$\Pe\to\tauh$ misid.                     &         \NA           &        \NA         &         40\%            &        \NA         \\
\cPqb tagging efficiency                 &      $<$6.5\%         &     $<$6.5\%       &      $<$6.5\%           &      $<$6.5\%      \\[1.0ex]
Embedded bkg.                            &         4\%           &        4\%         &         4\%             &        4\%         \\
\Zmm, \Pe{}\Pe bkg.                      &   4\% $\oplus$ 5\%    &  4\% $\oplus$ 5\%  &   4\% $\oplus$ 5\%      &  4\% $\oplus$ 5\%  \\
EW bkg.                                  &   4\% $\oplus$ 5\%    &  4\% $\oplus$ 5\%  &   4\% $\oplus$ 5\%      &  4\% $\oplus$ 5\%  \\
\wjets bkg.                              &         \NA           &        10\%        &         \NA             &        10\%        \\
Diboson bkg.                             &   5\% $\oplus$ 5\%    &  5\% $\oplus$ 5\%  &   5\% $\oplus$ 5\%      &  5\% $\oplus$ 5\%  \\
\ttbar bkg.                              &   6\% $\oplus$ 5\%    &  6\% $\oplus$ 5\%  &   6\% $\oplus$ 5\%      &  6\% $\oplus$ 5\%  \\
Single top quark bkg.                    &   5\% $\oplus$ 5\%    &  5\% $\oplus$ 5\%  &   5\% $\oplus$ 5\%      &  5\% $\oplus$ 5\%  \\
$\text{Jet}\to\tauh$ bkg.                &  30\% $\oplus$ 10\%   &        \NA         &  30\% $\oplus$ 10\%     &        \NA         \\[1.0ex]
Jet energy scale                         &        3--20\%        &       3--20\%      &        3--20\%          &       3--20\%      \\
\tauh energy scale                       &      0.7--1.2\%       &        \NA         &      0.7--1.2\%         &        \NA         \\
$\Pe\to\tauh$ energy scale               &        1--7\%         &        \NA         &        1--7\%           &        \NA         \\
$\PGm\to\tauh$ energy scale              &         1\%           &        \NA         &         1\%             &        \NA         \\
Electron energy scale                    &         \NA           &       1--2.5\%     &       1--2.5\%          &       1--2.5\%     \\
Muon energy scale                        &      0.4--2.7\%       &     0.4--2.7\%     &         \NA             &     0.4--2.7\%     \\[1.0ex]
Trigger timing inefficiency              &      0.2--1.3\%       &     0.2--1.3\%     &      0.2--1.3\%         &     0.2--1.3\%     \\
Integrated luminosity                    &         1.8\%         &        1.8\%       &         1.8\%           &        1.8\%       \\[1.0ex]
QCD scales (\Pg{}\Pg{}\PH)               &                                  \multicolumn{4}{c}{3.9\%}                                \\
QCD scales (VBF)                         &                                  \multicolumn{4}{c}{0.5\%}                                \\
PDF + \alpS (\Pg{}\Pg{}\PH)              &                                  \multicolumn{4}{c}{3.2\%}                                \\
PDF + \alpS (VBF)                        &                                  \multicolumn{4}{c}{2.1\%}                                \\
QCD acceptance (\Pg{}\Pg{}\PH)           &                           \multicolumn{4}{c}{$-$10.3 to +5.9\%}                           \\
QCD acceptance (VBF)                     &                           \multicolumn{4}{c}{$-$2.7 to +2.3\%}                            \\
PDF + \alpS acceptance (\Pg{}\Pg{}\PH)   &                           \multicolumn{4}{c}{$-$0.8 to +2.8\%}                            \\
PDF + \alpS acceptance (VBF)             &                           \multicolumn{4}{c}{$-$1.7 to +2.3\%}                            \\
\end{scotch}
}
\label{tab:systematics}
\end{table*}

\section{Results}
\label{sec:results}

No significant excess has been found for the LFV Higgs boson decays in both channels, and upper limits have been placed. Upper limits on the branching fraction of Higgs boson decay are computed using the modified frequentist approach for \CLs, taking the profile likelihood as a test statistic~\cite{Junk:1999kv, Read:2002hq, Cowan:2010js} in the asymptotic approximation. The observed (expected) upper limits on the Higgs boson branching fractions are 0.15 (0.15)\% for \Hmt and 0.22 (0.16)\% for \Het, respectively, at the 95\% \CL. The results have a dominant contribution from systematic uncertainties. The bin-by-bin uncertainties and the uncertainties related to the distribution of the misidentified lepton background have a significant impact followed by the lepton energy scale uncertainties.

\begin{figure*}
  \begin{center}
  \includegraphics[width=0.45\textwidth]{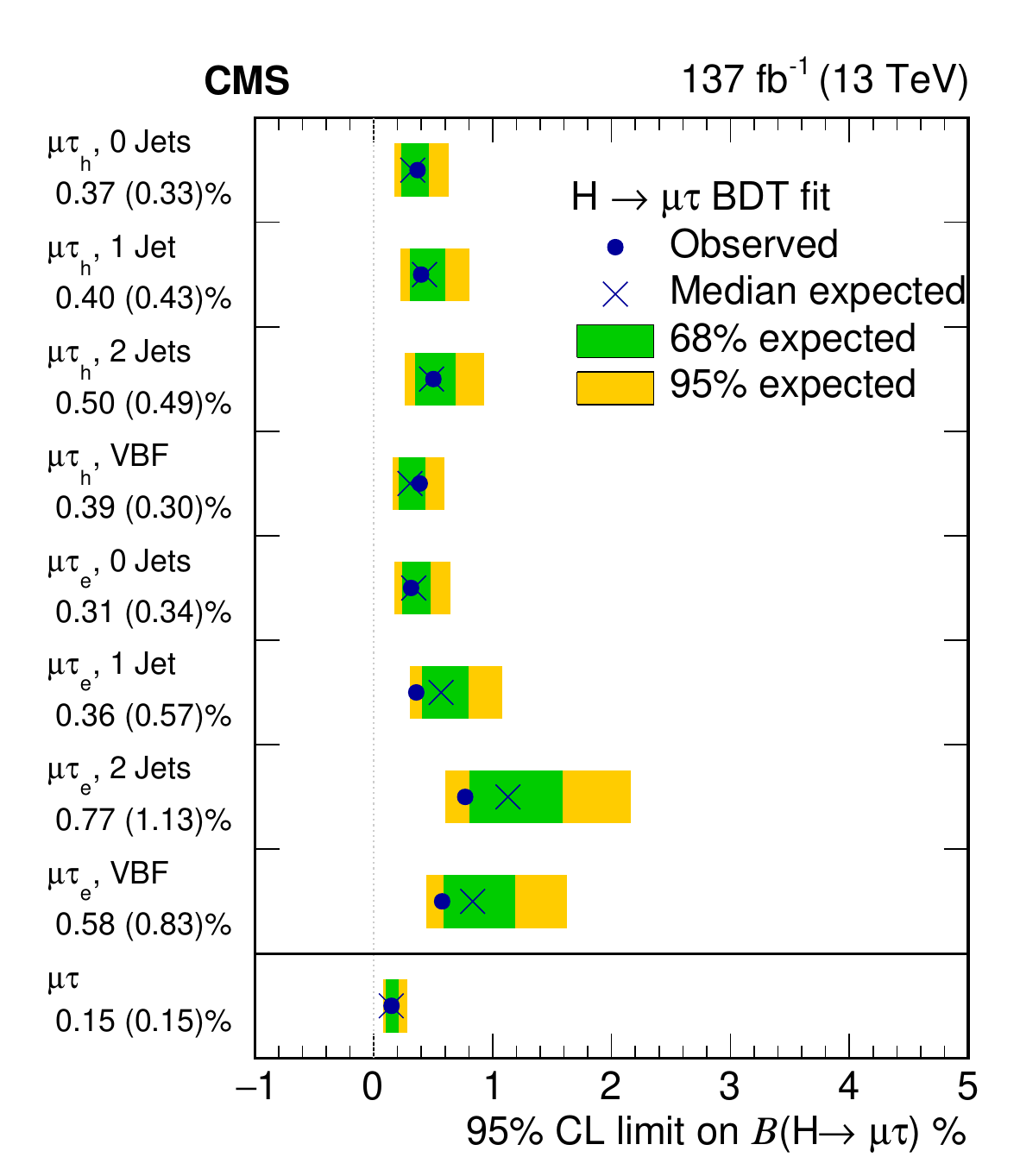}
  \includegraphics[width=0.45\textwidth]{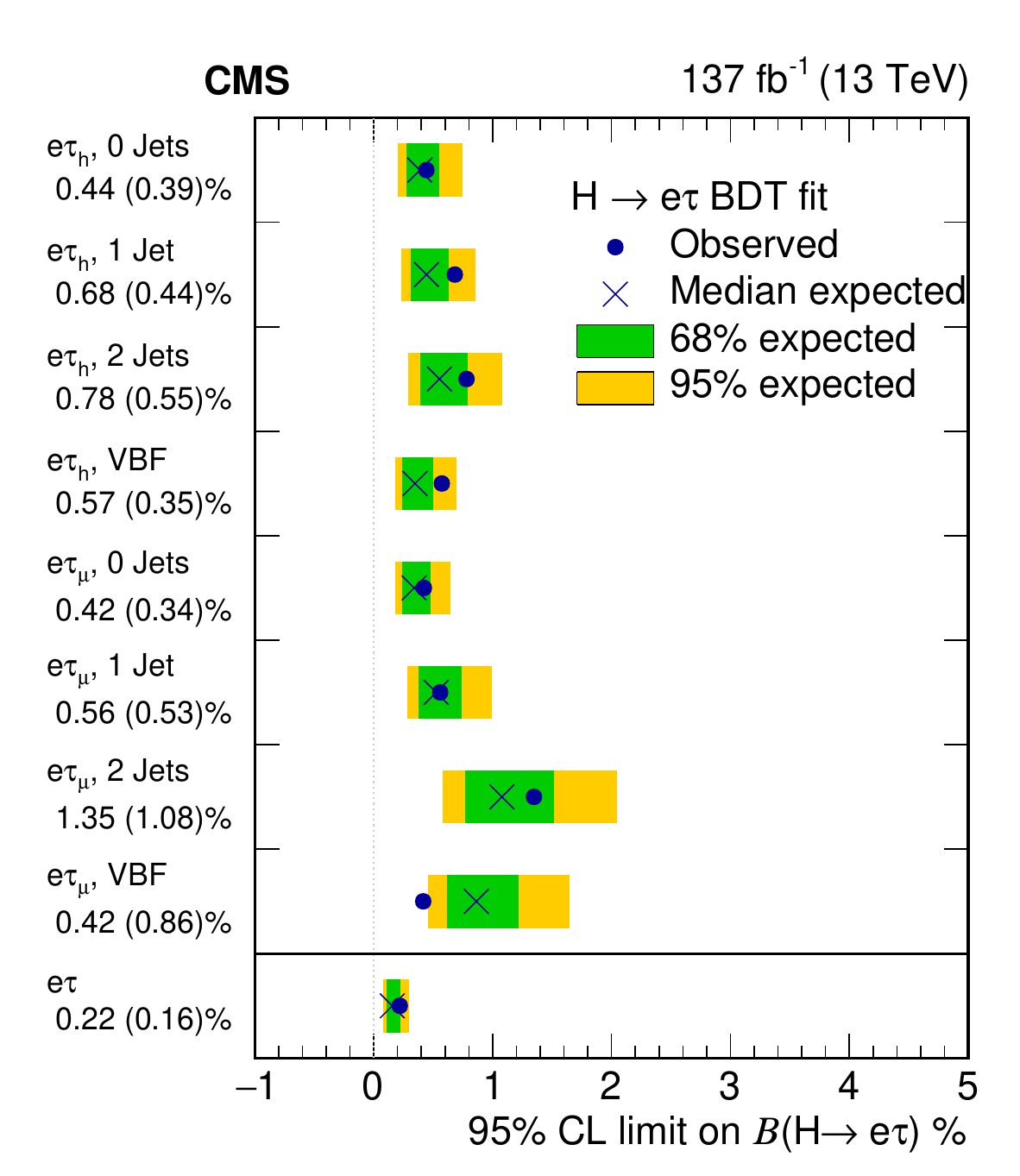} \\
  \caption{Observed (expected) 95\% \CL upper limits on the \BHmt (left) and \BHet (right) for each individual category and combined. The categories from top to bottom row are \muhad 0Jets, \muhad 1Jet, \muhad 2 Jets, \muhad VBF, \mue 0Jets, \mue 1Jet, \mue 2 Jets, \mue VBF, and \mutau combined (left) and \ehad 0Jets, \ehad 1Jet, \ehad 2 Jets, \ehad VBF, \emu 0Jets, \emu 1Jet, \emu 2 Jets, \emu VBF, and \etau combined (right).}
  \label{fig:bdt_limits}
  \end{center}
\end{figure*}

The upper limits and the best fit branching fractions for \BHmt and \BHet are reported in Tables~\ref{tab:limit_bdt_mutau} and~\ref{tab:limit_bdt_etau}. The limits are also summarized in Table~\ref{tab:limits_summary} and graphically shown in Fig.~\ref{fig:bdt_limits}. The limits are improved from previous results~\cite{Sirunyan:2017xzt}. The improvement relies on the larger data set, the updated background estimation techniques, and BDT classification. The results are cross-checked with an additional investigation following the strategy in Ref.~\cite{Sirunyan:2017xzt} and are found to be consistent.

The upper limits on \BHmt and \BHet are subsequently used to put constraints on LFV Yukawa couplings~\cite{Harnik:2012pb}. The LFV decays \Pe{}\PGt and \PGm{}\PGt arise at tree level from the assumed flavor violating Yukawa interactions, $Y_{\ell^\alpha\ell^{\beta}}$, where $\ell^\alpha,\ell^\beta$ are the leptons of different flavors ($\ell^{\alpha}\ne\ell^{\beta}$). The decay widths $\Gamma(\PH\to\ell^{\alpha}\ell^{\beta})$ in terms of the Yukawa couplings are given by:
\begin{linenomath*}
  \begin{equation*}
    \Gamma(\PH \to \ell^{\alpha} \ell^{\beta}) = \frac{\mh}{8\pi}(\abs{Y_{\ell^{\alpha}\ell^{\beta}}}^2 + \abs{Y_{\ell^{\beta}\ell^{\alpha}}}^2),
  \end{equation*}
\end{linenomath*}
and the branching fractions are given by:
\begin{linenomath*}
  \begin{equation*}
    \mathcal{B}(\PH \to \ell^{\alpha} \ell^{\beta}) = \frac{\Gamma(\PH \to \ell^{\alpha} \ell^{\beta})}{\Gamma(\PH \to \ell^{\alpha} \ell^{\beta}) + \Gamma_{\mathrm{SM}}}.
  \end{equation*}
\end{linenomath*}

\begin{figure*}
  \begin{center}
  \includegraphics[width=0.45\textwidth]{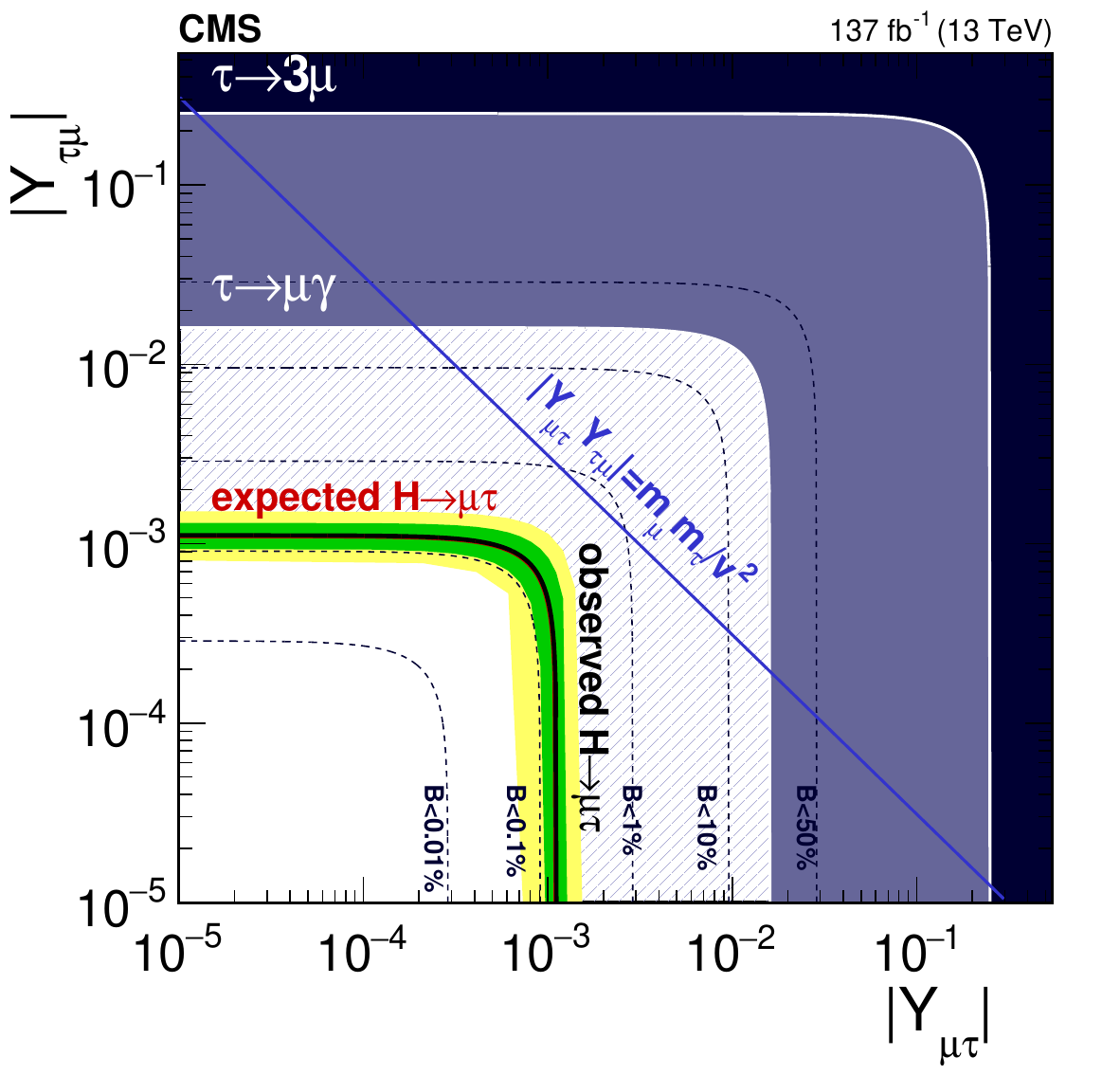}
  \includegraphics[width=0.45\textwidth]{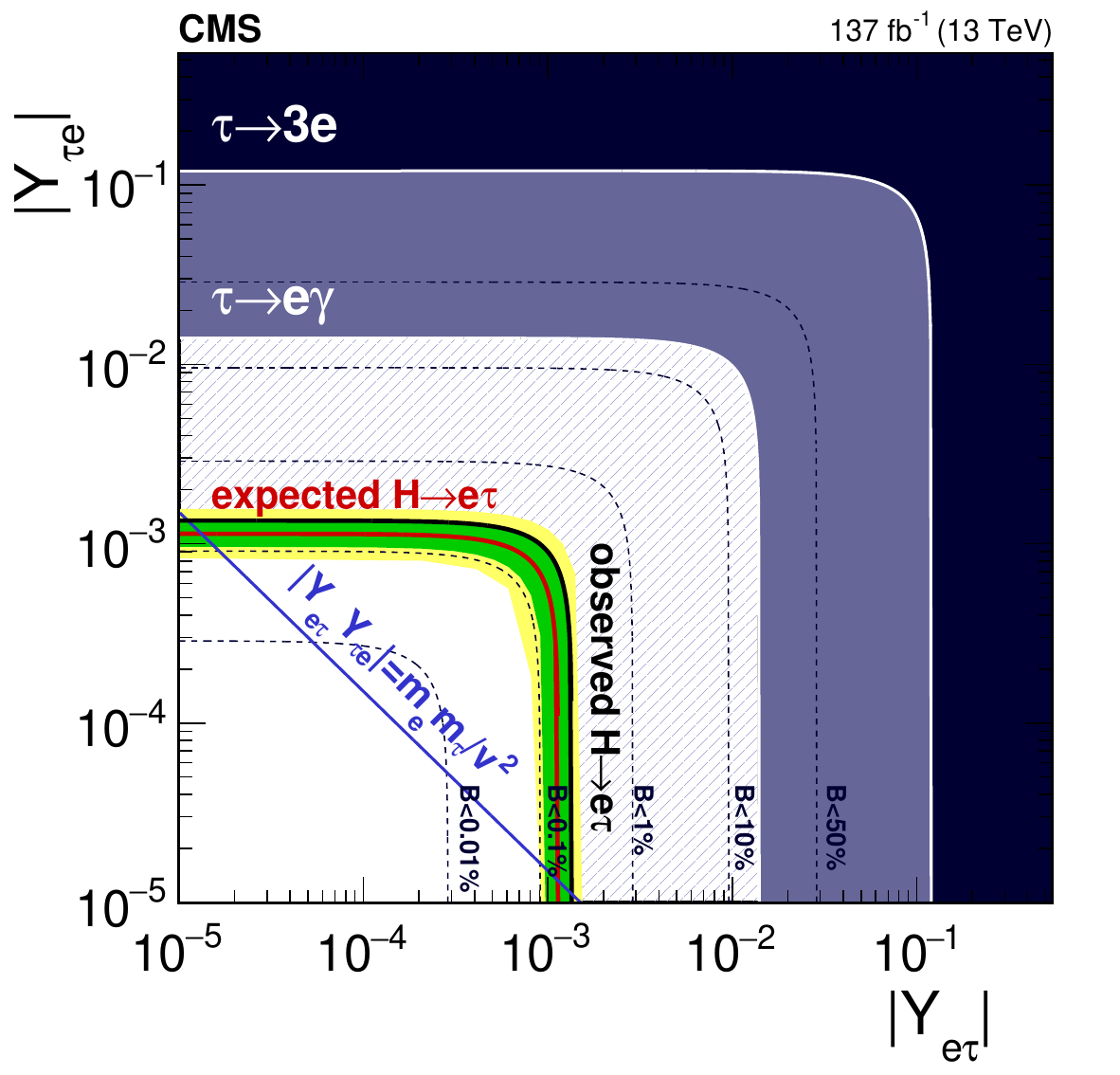} \\
  \caption{Expected (red line) and observed (black solid line) 95\% \CL upper limits on the LFV Yukawa couplings, $\Ymutau$ vs. $\Ytaumu$ (left) and $\Yetau$ vs. $\Ytaue$ (right). The $\Ymutau$ or $\Yetau$ couplings correspond to left chiral muon or electron and right chiral \PGt lepton, while $\Ytaumu$ or $\Ytaue$ couplings correspond to left chiral \PGt lepton and right chiral muon or electron. In the left plot, the expected limit is covered by the observed limit as they have similar values. The flavor diagonal Yukawa couplings are approximated by their SM values. The green and yellow bands indicate the range that is expected to contain 68\% and 95\% of all observed limit variations from the expected limit. The shaded regions are constraints obtained from null searches for $\PGt\to3\PGm$ or $\PGt\to3\Pe$ (dark blue)~\cite{Hayasaka:2010np} and $\PGt\to\PGm\Pgg$ or $\PGt\to\Pe\Pgg$ (purple)~\cite{Hayasaka:2007vc}. The blue diagonal line is the theoretical naturalness limit $\abs{Y_{ij}Y_{ji}} = {m_i}m_j/v^2$~\cite{Harnik:2012pb}.}
  \label{fig:bdt_yukawa_limits}
  \end{center}
\end{figure*}

The SM Higgs boson decay width is assumed to be $\Gamma_{\mathrm{SM}}=4.1\MeV$~\cite{Denner:2011mq} for $\mh=125\GeV$. The 95\% \CL upper limit on the Yukawa couplings obtained from the expression for the branching fraction above is shown in Table~\ref{tab:limits_summary}. The limits on the Yukawa couplings are $\sqrt{\smash[b]{\Ymutau^{2}+\Ytaumu^{2}}}<1.11{\times}10^{-3}$ and $\sqrt{\smash[b]{\Yetau^{2}+\Ytaue^{2}}}<1.35{\times}10^{-3}$ and are shown in Fig.~\ref{fig:bdt_yukawa_limits}. Tabulated results are available in the HepData database~\cite{hepdata}.

\begin{table*}[htbp]
\topcaption{Observed and expected upper limits at 95\% \CL and best fit branching fractions for each individual jet category, and their combinations, in the \Hmt channel.}
\begin{center}
\begin{scotch}{cccccc}
\multicolumn{6}{c}{Expected limits (\%)}                  \\
       & 0-jet   & 1-jet   & 2-jets  & VBF     & Combined \\
\mue   & $<$0.34 & $<$0.57 & $<$1.13 & $<$0.83 & $<$0.27  \\
\muhad & $<$0.33 & $<$0.43 & $<$0.49 & $<$0.30 & $<$0.18  \\
\mutau &         &         &         &         & $<$0.15  \\[1.0ex]
\multicolumn{6}{c}{Observed limits (\%)}                  \\
       & 0-jet   & 1-jet   & 2-jets  & VBF     & Combined \\
\mue   & $<$0.31 & $<$0.36 & $<$0.77 & $<$0.58 & $<$0.19  \\
\muhad & $<$0.37 & $<$0.40 & $<$0.50 & $<$0.39 & $<$0.24  \\
\mutau &         &         &         &         & $<$0.15  \\[1.0ex]
\multicolumn{6}{c}{Best fit branching fractions (\%)}                                       \\
       & 0-jet          & 1-jet          & 2-jets         & VBF            & Combined       \\
\mue   & $-0.03\pm0.17$ & $-0.40\pm0.28$ & $-0.66\pm0.56$ & $-0.41\pm0.39$ & $-0.14\pm0.13$ \\
\muhad & $+0.05\pm0.17$ & $-0.05\pm0.22$ & $+0.02\pm0.25$ & $+0.10\pm0.16$ & $+0.07\pm0.09$ \\
\mutau &                &                &                &                & $+0.00\pm0.07$ \\
\end{scotch}
\label{tab:limit_bdt_mutau}
\end{center}
\end{table*}

\begin{table*}[htbp]
\topcaption{Observed and expected upper limits at 95\% \CL and best fit branching fractions for each individual jet category, and their combinations, in the \Het channel.}
\begin{center}
\begin{scotch}{cccccc}
\multicolumn{6}{c}{Expected limits (\%)}                 \\
      & 0-jet     & 1-jet & 2-jets  & VBF     & Combined \\
\emu  & $<$0.34 & $<$0.53 & $<$1.08 & $<$0.86 & $<$0.26  \\
\ehad & $<$0.39 & $<$0.44 & $<$0.55 & $<$0.35 & $<$0.20  \\
\etau &         &         &         &         & $<$0.16  \\[1.0ex]
\multicolumn{6}{c}{Observed limits (\%)}                 \\
      & 0-jet   & 1-jet   & 2-jets  & VBF     & Combined \\
\emu  & $<$0.42 & $<$0.56 & $<$1.35 & $<$0.42 & $<$0.22  \\
\ehad & $<$0.44 & $<$0.68 & $<$0.78 & $<$0.57 & $<$0.37  \\
\etau &         &         &         &         & $<$0.22  \\[1.0ex]
\multicolumn{6}{c}{Best fit branching fractions (\%)}                                      \\
      & 0-jet          & 1-jet          & 2-jets         & VBF            & Combined       \\
\emu  & $+0.11\pm0.17$ & $+0.04\pm0.27$ & $+0.35\pm0.55$ & $-1.04\pm0.44$ & $-0.07\pm0.13$ \\
\ehad & $+0.07\pm0.20$ & $+0.29\pm0.23$ & $+0.27\pm0.29$ & $+0.27\pm0.17$ & $+0.20\pm0.10$ \\
\etau &                &                &                &                & $+0.08\pm0.08$ \\
\end{scotch}
\label{tab:limit_bdt_etau}
\end{center}
\end{table*}

\begin{table*}[htbp]
\topcaption{Summary of observed and expected upper limits at 95\% \CL, best fit branching fractions and corresponding constraints on Yukawa couplings for the \Hmt and \Het channels.}
\begin{center}
\begin{scotch}{lccc}
     & Observed (expected) & Best fit branching & Yukawa coupling                \\
     &  upper limits (\%)  &   fractions (\%)   &   constraints                  \\
\Hmt &   $<$0.15 (0.15)    &   $0.00\pm0.07$    & $<1.11\,(1.10){\times}10^{-3}$ \\
\Het &   $<$0.22 (0.16)    &   $0.08\pm0.08$    & $<1.35\,(1.14){\times}10^{-3}$ \\
\end{scotch}
\label{tab:limits_summary}
\end{center}
\end{table*}

\section{Summary}
\label{sec:summary}

A search for lepton-flavor violation has been performed in the \PGm{}\PGt and \Pe{}\PGt final states of the Higgs boson in data collected by the CMS experiment. The data correspond to an integrated luminosity of 137\fbinv of proton-proton collisions at a center-of-mass energy of 13\TeV. The results are extracted through a maximum likelihood fit to a boosted decision tree output, trained to distinguish the expected signal from backgrounds. The observed (expected) upper limits on the branching fraction of the Higgs boson to \PGm{}\PGt are 0.15 (0.15)\% and to \Pe{}\PGt are 0.22 (0.16)\%, respectively, at 95\% confidence level. Upper limits on the off-diagonal \PGm{}\PGt and \Pe{}\PGt couplings are derived from these constraints, $\sqrt{\smash[b]{\Ymutau^{2}+\Ytaumu^{2}}}<1.11{\times}10^{-3}$ and $\sqrt{\smash[b]{\Yetau^{2}+\Ytaue^{2}}}<1.35{\times}10^{-3}$. These results constitute an improvement over the previous limits from CMS and ATLAS experiments.

\begin{acknowledgments}
  We congratulate our colleagues in the CERN accelerator departments for the excellent performance of the LHC and thank the technical and administrative staffs at CERN and at other CMS institutes for their contributions to the success of the CMS effort. In addition, we gratefully acknowledge the computing centers and personnel of the Worldwide LHC Computing Grid and other centers for delivering so effectively the computing infrastructure essential to our analyses. Finally, we acknowledge the enduring support for the construction and operation of the LHC, the CMS detector, and the supporting computing infrastructure provided by the following funding agencies: BMBWF and FWF (Austria); FNRS and FWO (Belgium); CNPq, CAPES, FAPERJ, FAPERGS, and FAPESP (Brazil); MES (Bulgaria); CERN; CAS, MoST, and NSFC (China); MINCIENCIAS (Colombia); MSES and CSF (Croatia); RIF (Cyprus); SENESCYT (Ecuador); MoER, ERC PUT and ERDF (Estonia); Academy of Finland, MEC, and HIP (Finland); CEA and CNRS/IN2P3 (France); BMBF, DFG, and HGF (Germany); GSRT (Greece); NKFIA (Hungary); DAE and DST (India); IPM (Iran); SFI (Ireland); INFN (Italy); MSIP and NRF (Republic of Korea); MES (Latvia); LAS (Lithuania); MOE and UM (Malaysia); BUAP, CINVESTAV, CONACYT, LNS, SEP, and UASLP-FAI (Mexico); MOS (Montenegro); MBIE (New Zealand); PAEC (Pakistan); MSHE and NSC (Poland); FCT (Portugal); JINR (Dubna); MON, RosAtom, RAS, RFBR, and NRC KI (Russia); MESTD (Serbia); SEIDI, CPAN, PCTI, and FEDER (Spain); MOSTR (Sri Lanka); Swiss Funding Agencies (Switzerland); MST (Taipei); ThEPCenter, IPST, STAR, and NSTDA (Thailand); TUBITAK and TAEK (Turkey); NASU (Ukraine); STFC (United Kingdom); DOE and NSF (USA).

  \hyphenation{Rachada-pisek} Individuals have received support from the Marie-Curie program and the European Research Council and Horizon 2020 Grant, contract Nos.\ 675440, 724704, 752730, 765710 and 824093 (European Union); the Leventis Foundation; the Alfred P.\ Sloan Foundation; the Alexander von Humboldt Foundation; the Belgian Federal Science Policy Office; the Fonds pour la Formation \`a la Recherche dans l'Industrie et dans l'Agriculture (FRIA-Belgium); the Agentschap voor Innovatie door Wetenschap en Technologie (IWT-Belgium); the F.R.S.-FNRS and FWO (Belgium) under the ``Excellence of Science -- EOS" -- be.h project n.\ 30820817; the Beijing Municipal Science \& Technology Commission, No. Z191100007219010; the Ministry of Education, Youth and Sports (MEYS) of the Czech Republic; the Deutsche Forschungsgemeinschaft (DFG), under Germany's Excellence Strategy -- EXC 2121 ``Quantum Universe" -- 390833306, and under project number 400140256 - GRK2497; the Lend\"ulet (``Momentum") Program and the J\'anos Bolyai Research Scholarship of the Hungarian Academy of Sciences, the New National Excellence Program \'UNKP, the NKFIA research grants 123842, 123959, 124845, 124850, 125105, 128713, 128786, and 129058 (Hungary); the Council of Science and Industrial Research, India; the Ministry of Science and Higher Education and the National Science Center, contracts Opus 2014/15/B/ST2/03998 and 2015/19/B/ST2/02861 (Poland); the National Priorities Research Program by Qatar National Research Fund; the Ministry of Science and Higher Education, project no. 0723-2020-0041 (Russia); the Programa Estatal de Fomento de la Investigaci{\'o}n Cient{\'i}fica y T{\'e}cnica de Excelencia Mar\'{\i}a de Maeztu, grant MDM-2015-0509 and the Programa Severo Ochoa del Principado de Asturias; the Thalis and Aristeia programs cofinanced by EU-ESF and the Greek NSRF; the Rachadapisek Sompot Fund for Postdoctoral Fellowship, Chulalongkorn University and the Chulalongkorn Academic into Its 2nd Century Project Advancement Project (Thailand); the Kavli Foundation; the Nvidia Corporation; the SuperMicro Corporation; the Welch Foundation, contract C-1845; and the Weston Havens Foundation (USA).\end{acknowledgments}

\bibliography{auto_generated}
\cleardoublepage \appendix\section{The CMS Collaboration \label{app:collab}}\begin{sloppypar}\hyphenpenalty=5000\widowpenalty=500\clubpenalty=5000\vskip\cmsinstskip
\textbf{Yerevan Physics Institute, Yerevan, Armenia}\\*[0pt]
A.M.~Sirunyan$^{\textrm{\dag}}$, A.~Tumasyan
\vskip\cmsinstskip
\textbf{Institut f\"{u}r Hochenergiephysik, Wien, Austria}\\*[0pt]
W.~Adam, J.W.~Andrejkovic, T.~Bergauer, S.~Chatterjee, M.~Dragicevic, A.~Escalante~Del~Valle, R.~Fr\"{u}hwirth\cmsAuthorMark{1}, M.~Jeitler\cmsAuthorMark{1}, N.~Krammer, L.~Lechner, D.~Liko, I.~Mikulec, P.~Paulitsch, F.M.~Pitters, J.~Schieck\cmsAuthorMark{1}, R.~Sch\"{o}fbeck, M.~Spanring, S.~Templ, W.~Waltenberger, C.-E.~Wulz\cmsAuthorMark{1}
% ...
\vskip\cmsinstskip
\dag: Deceased\\
1:  Also at TU Wien, Wien, Austria\\
% ...
\end{sloppypar}
%%% END EDITABLE REGION %%%
% skeleton_end
\end{document}